\documentclass[manuscript]{acmart}

\AtBeginDocument{%
  }

\setcopyright{none}
\renewcommand\footnotetextcopyrightpermission[1]{}

\usepackage{multirow}
\usepackage{threeparttable}      
\usepackage{siunitx}             
\usepackage{subcaption}
\usepackage{mathtools}
\usepackage[capitalize,noabbrev]{cleveref}   

\usepackage[most]{tcolorbox}

\newtcolorbox{keytakeaway}{
  colback=black!3,
  colframe=black!35,
  boxrule=0.4pt,
  arc=1pt,
  left=6pt,
  right=6pt,
  top=5pt,
  bottom=5pt,
  before skip=6pt,
  after skip=6pt
}

\newcommand{\snapdate}{August 24, 2026}
\newcommand{\ngames}{1{,}974}
\newcommand{\nchallenge}{8{,}447}
\newcommand{\noverturn}{4{,}525}

\newcommand{\npitches}{4{,}114{,}256}
\newcommand{\numpires}{141}

\begin{document}

\title{When the Strike Zone Becomes Algorithmic: Umpire Judgment and Player Challenge Decisions under AI Review}

\author{Kichang Lee}
\authornote{These authors contributed equally to this work.}
\email{kichang.lee@kaist.ac.kr}
\affiliation{%
  \institution{KAIST}
  \city{Daejeon}
  \country{Republic of Korea}}
\affiliation{%
  \institution{Yonsei University}
  \city{Seoul}
  \country{Republic of Korea}}

\author{Gyeongmin Han}
\authornotemark[1]
\email{sd061123@yonsei.ac.kr}
\affiliation{%
  \institution{Yonsei University}
  \city{Seoul}
  \country{Republic of Korea}}

\author{Sungmin Lee}
\email{i.am.sungmin.lee@kaist.ac.kr}
\affiliation{%
  \institution{KAIST}
  \city{Daejeon}
  \country{Republic of Korea}}
\affiliation{%
  \institution{Yonsei University}
  \city{Seoul}
  \country{Republic of Korea}}

\author{JeongGil Ko}
\authornote{Corresponding author.}
\email{jeonggil.ko@yonsei.ac.kr}
\affiliation{%
  \institution{Yonsei University}
  \city{Seoul}
  \country{Republic of Korea}}

\renewcommand{\shortauthors}{Lee et al.}
\newcommand{\jk}[1]{{\color{red}[JK: #1]}}

\begin{abstract}
The Automated Ball-Strike challenge system that Major League Baseball adopted in 2026 offers a distinctive setting for studying human AI interaction in which umpires make every ball and strike call, while players can selectively ask an automated system to publicly overturn those decisions. We analyze 4,114,256 called pitches from 2015 through 2026 and 8,447 challenges from the 2026 season to examine how algorithmic review reshapes umpire judgment and player behavior. We study where umpires placed the effective strike zone boundary, how consistently they applied that boundary, how they responded to overturned calls, and which calls players chose to challenge. In 2026, the effective called boundary shifted toward the automated strike zone beyond the trajectory observed in prior seasons, while the consistency of that boundary largely continued its existing trend. Following an overturned call, umpires temporarily adjusted subsequent decisions near the corrected boundary, although these effects did not consistently persist into the next game. Count dependent variation in calling remained, while differences associated with player status narrowed. Players, meanwhile, left many overturnable calls unchallenged and appeared to base challenge decisions more strongly on immediately observable evidence than on the precise geometry of the automated zone. Together, these findings show that selective AI review does more than correct individual errors. It reshapes human judgment, adaptation, and strategic behavior around an algorithmic authority.
\end{abstract}





\maketitle


\section{Introduction}
\label{sec:intro}

For more than a century, ball-strike calls in baseball relied almost entirely on human umpires' judgment. The introduction of the Automated Ball-Strike system (ABS) has begun to change this practice. In 2024, the Korea Baseball Organization (KBO) introduced ABS to automate ball-strike calls throughout the league\cite{lee2025analyzing}. Major League Baseball (MLB), in contrast, adopted a challenge-based ABS in 2026. The home-plate umpire makes the initial call, but a batter, pitcher, or catcher may immediately challenge it. ABS then uses camera-based pitch tracking to uphold or overturn the call. This arrangement preserves human responsibility for the initial judgment while giving an automated system final authority over challenged decisions.

From a human-computer interaction (HCI) perspective, this arrangement raises a question about how automated review shapes human judgment~\cite{dietvorst2015algorithm,
lee2004trust, logg2019algorithm, parasuraman1997humans}. Prior research has examined systems that advise humans while leaving the final decision to them~\cite{bansal2021does, grgic2019human, lai2022human, passi2022overreliance, lee2026jarvis}, as well as algorithmic management systems that assign tasks and evaluate workers' performance~\cite{jarrahi2018artificial, kellogg2020algorithms, lee2015working,  rosenblat2016algorithmic, schildt2017big}. Challenge-based ABS, however, introduces a distinctive setting in which one person makes a decision, another chooses whether to contest it, and a system delivers a binding ruling. Understanding this interaction requires examining both how decision-makers respond to automated review and how the people entitled to request that review exercise their authority.


Automated review may influence umpire judgment in two ways: (i) through the possibility of review before a challenge is made and (ii) through the corrective feedback that follows an overturned call. These effects may change how willing an umpire is to call a strike or how accurately the umpire separates pitches near the zone boundary. Signal detection theory provides a framework for separating shifts in decision criterion from changes in perceptual sensitivity~\cite{green1966signal, hautus2021detection}. Corrective feedback may also lead to immediate conscious adjustments or, through repeated exposure, gradual perceptual recalibration~\cite{robert1998goldstone,
taylor2014explicit, watanabe2015perceptual}. We therefore examine changes associated with the availability of automated review together with changes after individual overturns, without assuming that either pathway affects only criterion or sensitivity.

\begin{figure*}[t]
  \centering
  \includegraphics[width=\linewidth]{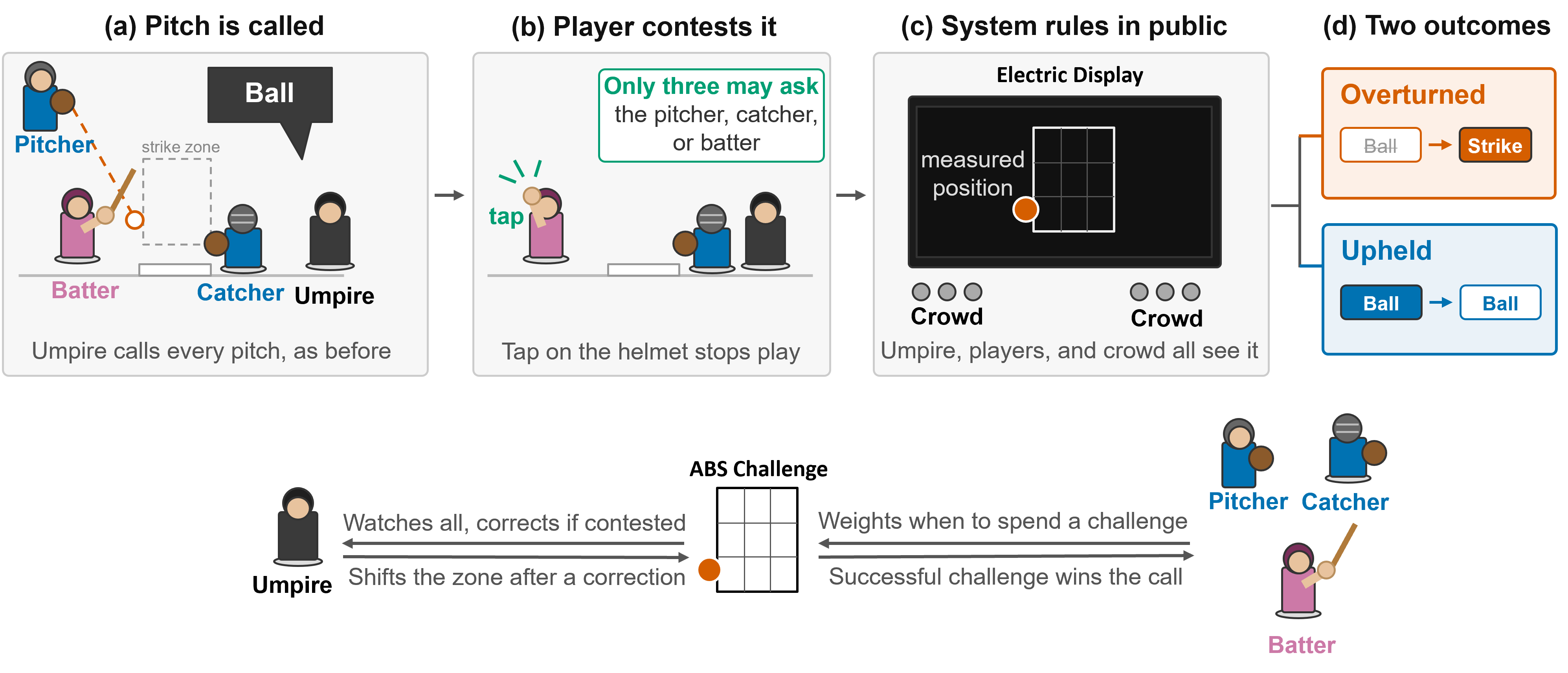}
  \caption{Overview of the MLB ABS challenge procedure. (a) The umpire makes the initial call; (b) the pitcher, catcher, or batter may challenge it; (c) ABS publicly displays the measured pitch location; and (d) the call is upheld or overturned. The lower schematic highlights the resulting asymmetry: umpires judge every pitch but receive correction only for challenged calls, while players determine which calls receive review.}
  \Description{Four panels illustrate the challenge procedure. Panel (a) shows an umpire calling a pitch a ball. Panel (b) shows a batter tapping their helmet to request review and identifies the pitcher, catcher, and batter as eligible challengers. Panel (c) shows a stadium display presenting the measured pitch location relative to the strike zone. Panel (d) illustrates the possible outcomes: the ball call is overturned to a strike or upheld as a ball. A lower schematic connects the umpire, the ABS system, and players to depict review, feedback, and challenge selection.}
  \label{fig:overview}
\end{figure*}

Figure~\ref{fig:overview} summarizes how an ABS challenge proceeds. Figure~\ref{fig:overview} (a) shows the umpire making the initial ball-strike call, after which the pitcher, catcher, or batter may contest it by tapping their helmet (see Figure~\ref{fig:overview} (b)). ABS then displays the measured pitch location (see Figure~\ref{fig:overview} (c)) and either upholds or overturns the call (c.f., Figure~\ref{fig:overview} (d)). This workflow creates two features central to our analysis. First, feedback is \emph{public and explicit}: an overturn directly reveals a discrepancy between the umpire's judgment and the system's ruling to the umpire, players, and spectators. Second, review is \emph{selective}: each team begins with two challenges and retains a challenge when successful, so players determine which eligible calls receive review. As summarized by the lower schematic, umpires judge every pitch but receive explicit correction only for calls that players choose to challenge. Across the \ngames{} games played from Opening Day through \snapdate{}, players initiated \nchallenge{} challenges (4.28 per game), with at least one challenge in 98.7\% of games; \noverturn{} (53.6\%) were overturned.

Prior research has studied systematic errors and contextual biases in human ball-strike judgments~\cite{deshpande2017hierarchical, hsu2024umpire, hunter2018new,
macmahon2008contextual, williams2019mlb, lee2026unified}, including associations with pitcher status~\cite{kim2014seeing, mills2014social, song2026technology}, the count~\cite{shinya2026bayesian}, and race~\cite{hamrick2015connection, parsons2011strike,
tainsky2015further}. Such patterns raise a distinction between perceptual limitations and the criteria used to translate perception into a call. Flannagan et al.~\cite{flannagan2024psychophysics} reported improvements in umpires' perceptual precision across seasons alongside persistent biases, suggesting that greater precision need not eliminate systematic decision tendencies. 
Evidence from tennis further suggests that automated review can accompany changes in officials' decision tendencies. Almog et al.~\cite{almog2024ai} reported a shift toward calling balls in after the introduction of Hawk-Eye. 

Taken together, however, prior work leaves several questions about challenge-based review unresolved. Baseball studies have documented broad patterns in umpire calls, and research on automated review has shown that judgments can change after such systems are introduced. These findings do not show how umpires respond to a specific overturn or whether any resulting adjustment persists beyond the immediate call. At the same time, studies of challenge behavior have generally treated the decision to request review as a separate problem, rather than as part of a feedback process that also involves the original decision-maker. Challenge-based ABS brings these elements together, allowing us to examine broader changes in umpire judgment, responses to individual overturns, and the calls players choose to challenge within the same system.


We investigate these questions using twelve seasons of MLB data from 2015 through 2026, comprising \npitches{} regular-season called pitches from \numpires{} home-plate umpires and \nchallenge{} challenge events recorded in 2026. Crucially, changes in 2026 must be interpreted with caution, since the introduction of challenge-based ABS also brought a new operational definition of the strike zone. A change in umpire calling patterns could therefore be associated with the new review system, the change in the zone used to evaluate pitches, or both.

To keep the change in strike-zone definition from being mistaken for a change in umpire judgment, we place all twelve seasons on the same spatial reference. We define a fixed reference zone based on batter height, independently of observed umpire calls, and account separately for how the strike zone used in each era differs from that reference. On this common scale, we estimate where umpires tend to switch from ball to strike at each edge of the zone and how sharply that switch occurs. These measures capture changes in strike-calling tendency and response precision, which we interpret in relation to decision criterion and sensitivity under our response model. We use them to examine both broader changes across seasons and changes following individual overturns, while analyzing players' challenge decisions separately.

Specifically, we focus on and address three research questions (RQ):

\begin{itemize}
  \item[\textbf{RQ1.}] How do umpires' judgments change after an individual overturn, in which direction and at which zone boundaries, and do these changes persist into the next game?
  \item[\textbf{RQ2.}] How do decision criteria, sensitivity, and condition-specific biases differ before and after the introduction of challenge-based ABS?
  \item[\textbf{RQ3.}] Which characteristics of a pitch and its game context are associated with players' decisions to challenge a call?
\end{itemize}


These questions frame challenge-based ABS as a system in which umpire judgment, automated feedback, and players' use of review are closely connected. By comparing broader changes across seasons with responses following individual overturns, we characterize how umpire calls vary under automated review across different timescales. From another perspective, we examine which calls players choose to challenge, since those choices determine when automated review and corrective feedback occur. Viewed together, these analyses show why understanding challenge-based review requires considering both how umpires respond to automated rulings and how players decide when to invoke them.

\section{Background and Related Work}
\label{sec:background}
\subsection{The ABS Challenge System}
\label{sec:background-abs}

Understanding the Automated Ball-Strike System (ABS) first requires understanding how ball-strike calls shape play. When a batter does not swing at a pitch, the home-plate umpire calls it a strike or a ball based on whether it passes through the strike zone. Each call updates the count, which records the number of balls and strikes accumulated during the current plate appearance. A fourth ball awards the batter first base, whereas a third strike results in a strikeout. A call therefore affects either the outcome of the plate appearance or the count under which play continues. The traditional rulebook defines the strike zone as a three-dimensional region above home plate, with vertical boundaries determined by the batter's stance. However, as Wang et al.~\cite{wang2026inside} explain, the written definition, umpires' calling practices, and technological representations of the zone have not always aligned.

MLB introduced challenge-based ABS in the 2026 season. Under this system, the umpire makes the initial call, and only the batter, pitcher, or catcher involved in the pitch may request review. Each team starts with two challenges and retains a challenge when it succeeds while losing one after an unsuccessful attempt. A team that enters an extra inning without a remaining challenge receives one. As illustrated in Figure~\ref{fig:overview}, a player must request review immediately after the call. The system then displays its ruling publicly before the call is upheld or overturned~\cite{wang2026inside}. As a result, a successful challenge changes the ruling while preserving the team's opportunity to request another review.

ABS applies a specific operational definition of the strike zone. Its zone is a 17-inch-wide rectangle at the midpoint of home plate, with its lower and upper boundaries set at 27\% and 53.5\% of the batter's measured standing height, respectively. Wang et al.~\cite{wang2026inside} trace how this implementation emerged through seven years of experimentation. Their study shows that both the geometry of the zone and the choice to use challenges were shaped by technical feasibility, economic considerations, continuity with established calling practices, and the preservation of gamesmanship. In particular, the challenge format preserves opportunities for players to influence and contest human judgments while making the decision to invoke automated review part of the game itself.

This arrangement distributes authority across three roles. The umpire makes the initial judgment, an eligible player decides whether to challenge it, and ABS determines the binding outcome of the review. Technological evaluation of umpires predates ABS through camera-based monitoring and public analyses of pitch-tracking data~\cite{bradbury2019monitoring, hunter2018new, mills2017technological, wang2026inside}. Challenge-based ABS adds immediate, player-initiated review that can reverse a call during the game. Umpire judgments are therefore subject to public correction, while whether that correction occurs depends on players' decisions to challenge.

\subsection{Automated Review and Human Authority}
\label{sec:background-authority}

Research on AI-assisted decision-making has examined how people accept or reject algorithmic advice~\cite{dzindolet2002perceived, green2019principles, poursabzi2021manipulating, yin2019understanding, zhang2020effect}. Bansal et al.~\cite{bansal2021does} found that explanations increased acceptance of AI recommendations regardless of their correctness, without further improving complementary team performance. Buccinca et al.~\cite{buccinca2021trust} showed that interventions encouraging deliberate engagement reduced overreliance compared with explanation-based interfaces. Together, these findings show that people's responses to AI depend partly on how they engage with algorithmic recommendations. Challenge-based ABS expands this into a distinctive settings where automated review can overturn the human decision rather than merely inform it. This shift in authority raises a related question about how people exercise judgment when an automated system can issue the final ruling.

Research on evaluation and accountability helps explain why such authority may influence judgment~\cite{frink2004advancing, hall2017accountability, ravid2020epm, ravid2023meta}. Rahman~\cite{rahman2021invisible} found that workers responded to opaque algorithmic evaluations by experimenting with ways to improve their scores or by restricting their platform activity. Their behavior changed even when the criteria behind the evaluation were difficult to understand. Lerner and Tetlock~\cite{lerner1999accounting} likewise show that accountability can reduce or amplify biases depending on the conditions under which people expect to justify their decisions. These findings motivate examining how judgment changes under evaluative authority without assuming that evaluation necessarily improves it.

Sports officiating provides evidence of these responses under technological monitoring and review. Bradbury~\cite{bradbury2019monitoring} examined QuesTec monitoring in baseball, finding some evidence of reduced shirking while emphasizing umpires' responsiveness to existing league directives. Studies of VAR in football similarly report partial reductions in home bias rather than its elimination~\cite{gasparetto2023does, holder2022monitoring, carlos2019does}. These findings establish that technological oversight can accompany uneven changes in officiating, although monitoring and human-led video review differ from systems that issue binding automated rulings.

Hawk-Eye in tennis provides a closer example of this form of review. More directly, Almog et al.~\cite{almog2024ai} found that Hawk-Eye review in tennis reduced overall errors on close calls while shifting judgments toward calling balls in. Their structural model interprets this shift in terms of the psychological cost of being overruled. The findings show that improved aggregate accuracy can coexist with a change in the kinds of errors officials make. They therefore motivate examining not only whether judgment becomes more accurate under automated review but also how calling patterns change.

Within baseball, Wang et al.~\cite{wang2026inside} explain how ABS emerged through negotiations over technological rule enforcement, preserving human participation while granting the system final authority over challenged calls. Their account of the system's design complements the behavioral evidence from other sports. Together, these perspectives motivate examining broader patterns in umpire judgment, responses following individual overturns, and the conditions under which players request review. This allows automated review to be studied through both the judgments subject to it and the human choices that determine when it occurs.

\subsection{Umpire Judgment: Bias, Feedback, and Measurement}
\label{sec:background-umpires}

Umpires are directly exposed to automated review given that their judgments are the decisions being evaluated and potentially overturned. Understanding their responses requires examining both the factors that shape their calls and the forms of change that may follow the corrective feedback. Prior research shows that ball-strike judgments depend on more than pitch location~\cite{chen2016decision, hsu2024umpire, lee2026auditing, macmahon2008contextual,
mills2014social, tainsky2015further}. Green and Daniels~\cite{green2014does} documented count-dependent variation in the called strike zone, while Kim and King~\cite{kim2014seeing} identified an association between pitcher status and favorable calls. Deshpande and Wyner~\cite{deshpande2017hierarchical} examined catcher framing, through which the presentation of a pitch influences its evaluation. These findings establish that the observed strike zone reflects contextual influences as well as the physical location of the ball.

Signal detection theory provides a framework for distinguishing two aspects of such judgments: (i) sensitivity, the ability to distinguish between alternatives, and (ii) criterion, the threshold used to select a response under uncertainty~\cite{green1966signal, hautus2021detection}. An umpire may become less willing to call a borderline pitch a strike without becoming better at distinguishing pitches inside and outside the zone. Conversely, improved discrimination need not eliminate a systematic tendency to favor one response. Flannagan et al.~\cite{flannagan2024psychophysics} illustrate this distinction by reporting improvements in perceptual precision across seasons alongside persistent biases. Their findings motivate examining the location and precision of the decision boundary separately, rather than treating changes in aggregate accuracy as a complete description of umpire behavior.

Corrective feedback introduces a further question: how does a particular overturn relate to subsequent judgments? Perceptual learning research describes how experience can alter the processing and differentiation of task-relevant information~\cite{fahle2005perceptual, robert1998goldstone, sagi2011perceptual, seitz2005unified,
watanabe2015perceptual}. Related work on sensorimotor adaptation distinguishes explicit adjustments from implicit learning~\cite{taylor2014explicit} and identifies adaptation processes operating at different rates~\cite{smith2006interacting}. These accounts motivate examining the time course and generalization of responses to correction. In baseball, a response may be concentrated near the corrected boundary, extend to other boundaries, or persist into later games. Such patterns provide evidence about the scope and duration of adjustment, although they do not uniquely identify an underlying cognitive mechanism.

Observing these changes requires an interpretable representation of the called strike zone. Existing approaches include geometric measures of accuracy and consistency~\cite{hunter2018new}, hierarchical models of contextual influences~\cite{deshpande2017hierarchical}, and psychophysical models that distinguish boundary location from the width of the transition between ball and strike responses~\cite{flannagan2024psychophysics}. Boundary location describes where the response changes, whereas transition width describes how gradually it changes with pitch position. Under the assumptions of the response model, these quantities support interpretations in terms of criterion and sensitivity. Longitudinal comparisons must additionally account for differences in the reference zone: Wang et al.~\cite{wang2026inside} show that the rulebook definition, established calling practices, and the zone implemented by ABS are not interchangeable. This distinction motivates using a common spatial reference across our twelve-season dataset while accounting separately for era-specific zone definitions.

\subsection{Player Decisions to Challenge}
\label{sec:background-players}

Players determine when automated review is invoked and, consequently, which umpire judgments receive public confirmation or correction. A challenge decision involves both uncertainty about the call and the value of requesting review in the current game situation~\cite{gigerenzer1996reasoning, tversky1974judgment}. Disagreement with the umpire may favor a challenge, but the possibility of losing a remaining opportunity for review may favor restraint. Understanding challenge behavior therefore requires considering perceptual judgment alongside the incentives created by the review procedure.

Research in tennis demonstrates the relevance of both factors. Mather~\cite{mather2008perceptual} analyzed line-call challenges using a psychophysical model of uncertainty in players' and officials' judgments. The model accounted for the concentration of challenges and errors near court boundaries, showing how perceptual uncertainty can explain observed challenge patterns. From a decision-theoretic perspective, Abramitzky et al.~\cite{abramitzky2012optimality} modeled challenges as a trade-off between the immediate benefit of overturning a call and the value of retaining opportunities for later use. They found that players' behavior was close to the optimal strategy prescribed by their model. Together, these studies suggest that challenge decisions reflect both whether a player believes a call is incorrect and whether contesting it is worthwhile.

In baseball, Wang et al.~\cite{wang2026inside} identify selective enforcement as a design choice that preserves gamesmanship and makes the use of automation part of player strategy. This perspective motivates examining how challenge requests vary with pitch characteristics, the count, and the broader game context. However, an unchallenged call does not reveal whether the player accepted the ruling, failed to recognize a discrepancy, or deliberately preserved a challenge. Similarly, the pitch location measured by ABS does not directly reveal what the player perceived. We therefore examine observable conditions associated with review requests without treating challenge rates alone as evidence of perceptual ability or strategic intent. These choices also define the set of calls on which umpires receive automated feedback, connecting players' use of the system to the study of umpire responses.

\subsection{Positioning the Present Study}
\label{sec:background-positioning}

Prior research has examined technological officiating as a broader class of decision-aid systems, including questions of authority, accuracy, and standards of review~\cite{collins2010philosophy,kolbinger2017scientific}, and documented how technological monitoring and review alter officials' decisions in baseball, football, and tennis~\cite{almog2024ai, bradbury2019monitoring,
holder2022monitoring, carlos2019does, mills2017technological, spitz2021video} and modeled challenge behavior as a combination of perceptual uncertainty and the strategic use of limited review opportunities~\cite{abramitzky2012optimality, mather2008perceptual,
whitney2008perceptual}. In MLB specifically, recent work has traced how ABS developed from a technological aid into a system of selective, binding rule enforcement, showing that the operational strike zone reflects design choices rather than a straightforward translation of the written rule~\cite{wang2026inside}.

Our study brings these perspectives together in MLB's challenge-based ABS by examining umpire judgments and players' requests for review within the same setting. Using twelve seasons of pitch data and challenge records from 2026, we characterize longer-term judgment patterns, the spatial specificity and persistence of changes following overturns, and the conditions associated with challenge requests. Building on psychophysical approaches that model the called strike zone in terms of boundary location and response precision~\cite{flannagan2024psychophysics}, we use a common spatial reference to compare judgments across seasons while separately accounting for changes in the operational definition of the zone~\cite{wang2026inside}. This approach examines both how judgments vary under automated review and how participants selectively invoke that review, without assuming that observed behavioral patterns uniquely identify cognitive mechanisms or strategic intent.

\section{Data and Measurement Framework}
\label{sec:data}

\subsection{Dataset Construction and Measurement Harmonization}
\label{sec:data-harmonization}

\vspace{1ex} \noindent \textbf{Sample construction.}
Our dataset draws on MLB pitch-tracking data from Baseball Savant covering twelve regular seasons from 2015 through 2026, with the 2026 data available through August 24. The analysis is restricted to balls and called strikes to focus on pitches that require a ball-strike decision by the home-plate umpire. Extra innings are excluded to keep the number of available challenge opportunities comparable across games, as MLB provides each team with an additional challenge in every extra inning. For each pitch, the dataset includes its tracked location, strike-zone boundaries, home-plate umpire, batter, and ball-strike count. After applying these criteria, the final sample contains \npitches{} called pitches, including 8,283 challenged pitches, adjudicated by \numpires{} home-plate umpires. We note that before comparing these pitches across seasons two changes in the 2026 data must be addressed.

\vspace{1ex} \noindent \textbf{Original umpire calls.}
The first change concerns the recorded outcome of challenged pitches. In 2026, this outcome reflects the final ruling after ABS review and may differ from the umpire's original call. For each overturned challenge, the recorded outcome is reversed to recover the call initially made by the umpire. We denote this original call by $y_i$, where $y_i=1$ indicates a called strike and $y_i=0$ indicates a called ball.

\vspace{1ex} \noindent \textbf{Common measurement plane.}
The second change concerns where pitch location is measured. Through 2025, pitch coordinates were reported at the front plane of home plate, whereas the 2026 season coordinates are reported at the midpoint plane used by ABS. As a result, the same pitch would have different reported coordinates depending on which plane is used for measurement. The size and direction of this difference depend on the trajectory of the pitch. Consequently, comparing the reported coordinates directly could make a change in measurement convention appear as a change in umpire judgment.

To make the coordinates comparable across seasons, we reconstruct each pre-2026 season pitch at the midpoint plane of home plate. For each pitch, velocity and acceleration parameters from the tracking data are used to determine when it reaches the front and midpoint planes. Propagating its trajectory between these two points yields the horizontal and vertical coordinates at the midpoint. Coordinates from 2026 are already reported at this plane and require no reconstruction. This procedure places pitch locations from all twelve seasons on the same measurement plane. We next define a common spatial reference on this plane so that changes in the operational strike-zone definition are not mistaken for changes in umpire judgment.

\subsection{Spatial Reference and Edge-Specific Samples}
\label{sec:data-frame}

Comparing umpire judgment across seasons requires a spatial reference that remains fixed over time. This is especially important in 2026, when challenge-based ABS was introduced together with a new operational definition of the strike zone. If each season were measured relative to its own operational zone, movement in the reference boundary could offset or exaggerate movement in the called boundary. For this, we measure every season relative to a single fixed reference and account separately for differences in the official strike-zone definition.

We use the following 2026 ABS zone geometry as a fixed reference for all seasons.
\begin{equation}
\label{eq:ruler}
R^{\mathrm{top}}_b = 0.535\,h_b, \qquad
R^{\mathrm{bot}}_b = 0.270\,h_b, \qquad
R^{\mathrm{side}} = W, \qquad
W = 8.5~\mathrm{in},
\end{equation}
where, $h_b$ is the height of batter $b$. The top and bottom edges vary with batter height, while the two side edges are located at $\pm W$. Applying this geometry to all twelve seasons provides a common measurement reference. Note that we use this geometry only as a common measurement reference and do not assume that it was the strike zone applied before 2026.

The choice of the 2026 geometry also reflects how the automated zone was developed. Wang et al.~\cite{wang2026inside} describe how earlier implementations based more directly on the three-dimensional rulebook zone produced calls that players found unfamiliar. MLB subsequently calibrated a two-dimensional zone, including its placement at the midpoint of home plate, to better align automated rulings with established calling practices~\cite{wang2026inside}. The resulting geometry differs from earlier rule definitions in physical space but was developed with reference to the ball-strike boundary experienced by players and umpires. It therefore provides a meaningful fixed reference without requiring us to treat it as the historical rule.

A pitch is ruled a strike when any part of the ball intersects the zone. Accordingly, we expand each geometric edge by the radius of the baseball, $\rho=1.45$ inches.
\begin{equation}
\tilde R^{\mathrm{top}}_b = R^{\mathrm{top}}_b + \rho, \qquad
\tilde R^{\mathrm{bot}}_b = R^{\mathrm{bot}}_b - \rho, \qquad
\tilde R^{\mathrm{side}} = R^{\mathrm{side}} + \rho.
\end{equation}

With the reference boundary defined, each pitch can be represented by its signed distance from a particular edge. We set $d=0$ at the reference boundary, with positive values toward the interior of the zone and negative values outside.
\begin{equation}
\label{eq:dist}
d^{\mathrm{top}}_i = \tilde R^{\mathrm{top}}_{b(i)} - z_i, \qquad
d^{\mathrm{bot}}_i = z_i - \tilde R^{\mathrm{bot}}_{b(i)}, \qquad
d^{\mathrm{ins}}_i = \tilde R^{\mathrm{side}} - s_i x_i, \qquad
d^{\mathrm{out}}_i = \tilde R^{\mathrm{side}} + s_i x_i,
\end{equation}
where $s_i=-1$ for right-handed hitters and $s_i=+1$ for left-handed hitters. This transformation gives the inside and outside edges the same orientation for both batter sides. Under this representation, $d_i^e=1$ means that the center of the ball lies one inch inside reference edge $e$, while $d_i^e=-1$ means that it lies one inch outside.

Pitches closest to an edge provide the most information about where calls change from ball to strike and how sharply that change occurs. We estimate each edge using pitches within $b=3.0$ inches of its reference boundary. Across the observed season-edge cells, this band spans 2.34 to 4.26 estimated transition widths $\sigma$ defined below, with a median of 2.95.

Pitches near a corner require an additional restriction because two boundaries may influence the call. For a target edge $e$, we retain pitches that lie within $b$ inches of that edge while remaining sufficiently far inside both orthogonal edges. The quantity $|d_i^e|$ gives the absolute distance from the target edge regardless of whether the pitch lies inside or outside the zone. For each orthogonal edge $o$, a larger positive $d_i^o$ indicates that the pitch lies farther inside that boundary. The resulting edge sample is
\begin{equation}
\label{eq:band}
B_e =
\left\{
i :
|d_i^e| \le b,
\quad
\min_{o \perp e} d_i^o > c
\right\}.
\end{equation}
The first condition keeps pitches near the target edge, while the second removes pitches close to corners where another boundary may also influence the call. For consistency across seasons and edges, we use a single corner threshold $c$. We choose this threshold so that pitches retained for a target edge have a fitted strike probability of at least 0.95 with respect to each orthogonal edge across all observed season-edge cells. A threshold of $c=3.7$ inches satisfies this requirement, while the corresponding cell-specific thresholds range from 1.13 to 3.66 inches. Applying these conditions leaves 842{,}456 pitches, which we refer to as the \emph{band sample}. The four edge samples are mutually disjoint, so each pitch is assigned to at most one edge and each edge is estimated from pitches for which it is the primary spatial constraint on the call.

\subsection{Estimating Criterion and Sensitivity}
\label{sec:data-metrics}

Within each edge sample, the two-dimensional strike zone reduces to a one-dimensional decision problem. Each pitch is represented by its signed distance $d_i^e$ from the reference edge and the umpire's original call $y_i$. As $d_i^e$ increases from outside the zone toward its interior, the probability of a strike call increases.

Following the psychophysical model of umpire judgment proposed by Flannagan et al.~\cite{flannagan2024psychophysics} and Lee et al.~\cite{lee2025analyzing}, we adapt their logistic formulation to our edge-specific signed-distance representation:
\begin{equation}
\label{eq:psychometric}
\Pr(y_i=1)
=
\operatorname{logit}^{-1}
\left(
\frac{d_i^e+\alpha}{\sigma}
\right).
\end{equation}
Whereas Flannagan et al.\ model pitch location using distance from the center of the strike zone, our formulation measures signed distance from each individual edge. This allows us to interpret the two parameters directly in terms of boundary judgment: $\alpha$ captures the location of the called boundary relative to the common reference, while $\sigma$ captures the width of the transition from ball to strike calls.

The two parameters describe different aspects of umpire judgment. The parameter $\alpha$ determines the location of the transition. The fitted strike probability equals 0.5 at $d_i^e=-\alpha$, so $\alpha>0$ indicates that the called boundary lies outside the common reference, while $\alpha<0$ indicates that it lies inside. We therefore interpret $\alpha$ as the outward displacement of the umpire's decision criterion, measured in inches.

The parameter $\sigma$ determines the width of the transition. Smaller values produce a steeper change from ball to strike calls around the boundary, while larger values produce a more gradual transition. For example, moving from $-\sigma$ to $+\sigma$ around the 50-percent point changes the fitted strike probability from approximately 0.27 to 0.73. Thus, smaller $\sigma$ indicates greater spatial sensitivity to changes in pitch location.

\begin{figure*}[t]
  \centering
  \includegraphics[width=\linewidth]{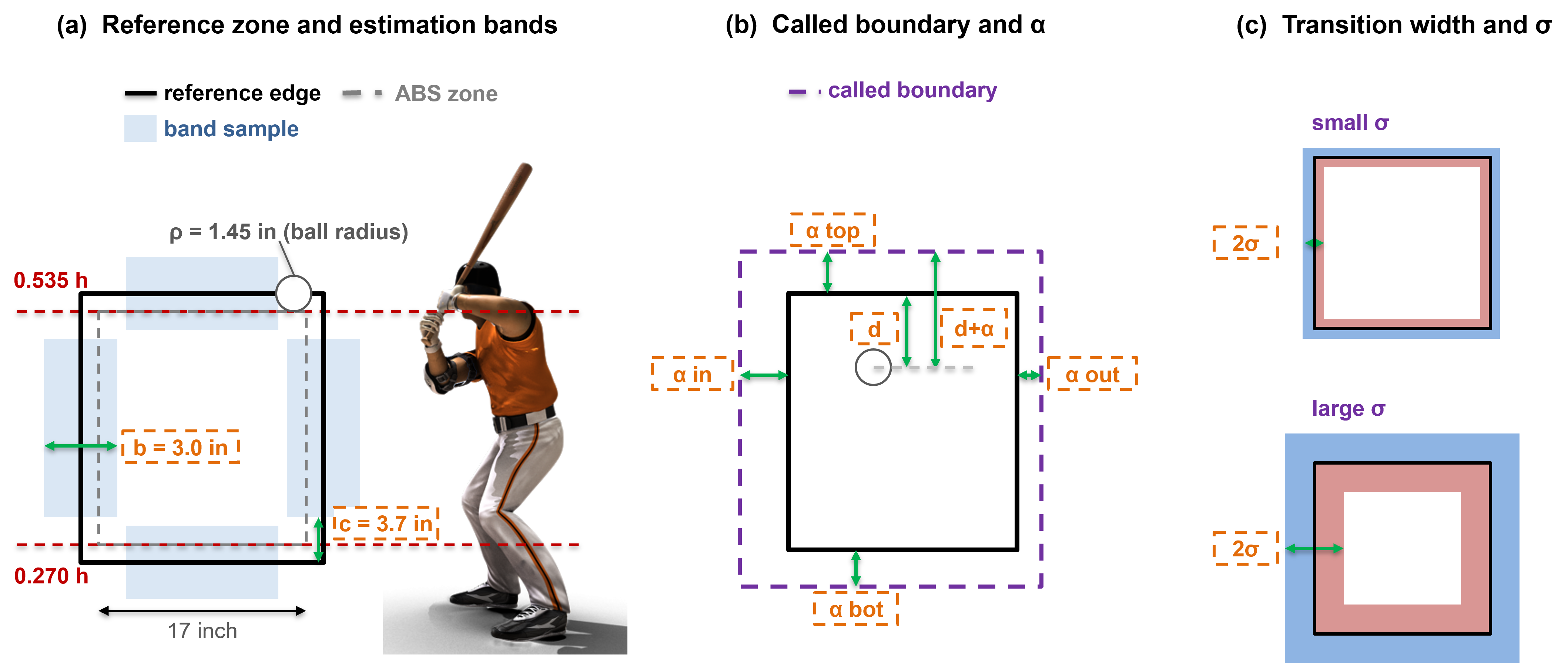}
  \caption{Illustration of the two quantities used to characterize umpire decision boundaries. (a) A fixed reference zone defines four disjoint edge bands used for estimation. (b) The per-edge displacement $\alpha$ measures how far the estimated called boundary lies from the reference edge. (c) The transition width $\sigma$ captures how sharply the boundary is applied: smaller $\sigma$ indicates a sharper transition, whereas larger $\sigma$ indicates a broader region in which similar pitch locations receive both calls.}
  \Description{Three panels. Panel (a) shows the reference zone as a solid black rectangle with the ABS zone as a grey dashed rectangle just inside it, and a batter in his stance beside it for scale. Red dashed lines mark the upper and lower edges at 0.535 and 0.270 times batter height, an arrow gives the plate width as 17 inches, and a circle at the upper corner marks the ball radius of 1.45 inches. Light blue shading marks the band sample along each of the four edges, labelled with the band half-width b and the corner threshold c, and the four bands do not meet at the corners. Panel (b) shows the same black reference rectangle with a purple dashed rectangle outside it for the called boundary. Green arrows between the two rectangles are labelled alpha top, alpha bottom, alpha inside, and alpha outside, one per edge, and a ball inside the zone is marked with its distance d to the reference edge and d plus alpha to the called boundary. Panel (c) stacks two diagrams of the same rectangle. The upper one, labelled small sigma, has a narrow pink band inside the edge and a narrow blue band outside it, with an arrow spanning them labelled two sigma. The lower one, labelled large sigma, has the same bands drawn much wider with a correspondingly longer two sigma arrow.}
  \label{fig:psychometric}
\end{figure*}

Figure~\ref{fig:psychometric} shows how $\alpha$ and $\sigma$ capture distinct aspects of umpire judgment. Figure~\ref{fig:psychometric} (a) shows the four disjoint edge bands used to estimate each boundary. Figure~\ref{fig:psychometric} (b) defines $\alpha$ as the displacement between the fixed reference edge and the estimated called boundary, capturing where the boundary is placed. Figure~\ref{fig:psychometric} (c) illustrates $\sigma$: a small $\sigma$ indicates a sharp and consistent transition between ball and strike calls, whereas a large $\sigma$ indicates greater ambiguity around the boundary. Thus, $\alpha$ measures \emph{where} the boundary lies, while $\sigma$ measures \emph{how sharply} it is applied.

\vspace{1ex} \noindent \textbf{Adjustment variables.}
Equation~\eqref{eq:psychometric} captures the core relationship between pitch location and the probability of a strike call. In the observed data, however, calls may also vary with the official strike-zone definition, batter geometry, game context, and other factors. If these differences are ignored, they could shift the fitted psychometric curve and be incorrectly attributed to changes in criterion or sensitivity. Given this, we add an adjustment term $A_i$:
\begin{equation}
\label{eq:adjusted-psychometric}
\Pr(y_i=1)
=
\operatorname{logit}^{-1}
\left(
\frac{d_i^e+\alpha}{\sigma}
+
A_i
\right).
\end{equation}

The adjustment term collects four types of factors that we want to account for separately from the spatial transition:
\begin{equation}
\label{eq:adjustments}
A_i
=
\lambda\,\Delta Z_{b(i)}
+
\rho_R\,\mathrm{Rc}_i
+
\boldsymbol{\gamma}^{\top}\tilde{\mathbf{X}}_i
+
\boldsymbol{\tau}^{\top}\mathbf{T}_i.
\end{equation}

The first term, $\Delta Z_{b(i)}$, accounts for differences between the official strike-zone edge applicable to batter $b(i)$ and the fixed spatial reference used in our analysis. Recall that we measure every season using the same reference geometry so that criterion is expressed on a common scale. However, the official zone itself can differ from this reference and can change across seasons. Including $\Delta Z_{b(i)}$ allows the model to account for these differences rather than absorbing them into the estimated umpire criterion.

For the top and bottom edges, we additionally include $\mathrm{Rc}_i$ to account for batter geometry. While our reference zone already scales with batter height, umpires may not adjust the vertical boundary perfectly in proportion to that geometry. $\mathrm{Rc}_i$ measures how far the batter-specific reference edge differs from the corresponding seasonal average. We center this variable within each season, so $\mathrm{Rc}_i=0$ represents the average batter geometry for that season.

The covariate block $\tilde{\mathbf{X}}_i$ accounts for other characteristics that may systematically affect ball-strike calls. Following the psychophysical specification of Flannagan et al.~\cite{flannagan2024psychophysics}, it includes eleven count indicators, three batter-by-pitcher handedness indicators, inning half, and umpire indicators. We center each covariate by its mean in the estimation sample. Consequently, $\tilde{\mathbf{X}}_i=\mathbf{0}$ represents the average composition of pitches in that sample rather than an arbitrary combination of reference categories.

Finally, $\mathbf{T}_i$ contains variables specific to the analysis being conducted. For example, later analyses may use these terms to distinguish pitches before and after a particular feedback event or between treatment and comparison conditions. We leave these variables uncentered so that $\mathbf{T}_i=\mathbf{0}$ corresponds to the relevant baseline condition. The specific treatment terms and their interpretation are introduced when each analysis is presented.

With these adjustments, $\alpha$ can be interpreted under a common reference condition in which the covariates and batter geometry are centered at their mean values, the official zone edge coincides with the fixed spatial reference, and the treatment is set to the baseline condition. Thus, $A_i$ captures systematic variation unrelated to the spatial judgment of interest, allowing $\alpha$ and $\sigma$ to characterize the location and sharpness of the umpire's ball strike transition.

\vspace{1ex} \noindent \textbf{Estimation and inference.}
For estimation, Equation~\eqref{eq:adjusted-psychometric} is fit in the equivalent linear logistic form
\begin{equation}
\label{eq:estimation-model}
\operatorname{logit}\Pr(y_i=1)
=
\beta_0
+
\beta_d d_i^e
+
A_i.
\end{equation}
The psychophysical parameters are then recovered as
\begin{equation}
\label{eq:metrics}
\alpha=\frac{\beta_0}{\beta_d},
\qquad
\sigma=\frac{1}{\beta_d}.
\end{equation}

We estimate the model separately for each season and edge using Newton's method. Because calls may be correlated within both umpires and batters, we use a two-way cluster sandwich estimator with clustering by umpire and batter. Standard errors ($\mathrm{se}$) for $\alpha$ and $\sigma$ are obtained using the delta method, and all reported regression-based intervals use $1.96\times\mathrm{se}$.

\subsection{Benchmarking 2026 Against the Pre-Adoption Trend}
\label{sec:data-trend}

A direct comparison between 2025 and 2026 is insufficient because criterion and sensitivity were already changing before challenge-based ABS was introduced. We therefore evaluate whether the 2026 estimates depart from the longer-term trajectory observed before adoption. For each edge-specific $\alpha$ and $\sigma$ series, we fit a linear trend using the 11 seasons from 2015 through 2025 and project that trend one year forward. We then compare the observed 2026 estimate with the value predicted from the pre-adoption trajectory.

We give the 11 annual observations equal weight. The residual year-to-year variation in these series is substantially larger than the estimation error of the individual seasonal estimates. Weighting seasons by their estimation precision would give seasons with larger within-season samples greater influence over the temporal slope even though they are not necessarily more representative of the underlying year-to-year trajectory. Our inferential target is a new annual observation rather than the mean fitted trend. Therefore, we evaluate the 2026 season data using the predictive distribution for a new observation from the pre-adoption regression, which incorporates both uncertainty in the fitted trend and residual year-to-year variation around it.

For each series, we report the percentile of the observed 2026 value within this predictive distribution together with its 50 and 95 percent prediction intervals. The 95 percent interval provides a conventional benchmark for identifying observations outside the range implied by the pre-adoption trajectory, while the 50 percent interval provides a narrower reference for the distance from its center. These comparisons indicate whether the 2026 criterion and sensitivity estimates continue the historical trajectory or depart from it following the introduction of challenge-based ABS.

\section{RQ1. How Do Umpire Calls Change Under Challenge-Based ABS?}
\label{sec:rq1}

\subsection{Analytical Overview}
\label{sec:rq1-question}

Challenge-based ABS creates two levels at which changes in umpire calls may appear. At the system level, every call becomes eligible for review once the challenge system is introduced, even when no challenge is made. At the event level, an overturn provides explicit feedback that a particular call disagreed with the automated ruling. We first examine the system level by asking whether the called boundary $\alpha$ or transition width $\sigma$ in 2026 departed from the 2015--2025 trajectory estimated in Section~\ref{sec:data-metrics}. Since all umpires entered the challenge system at the same time, this comparison identifies a departure from the pre-adoption pattern rather than a causal effect of reviewability.

We then turn to individual challenge rulings. Here, the question is more specific. ``After an umpire receives an overturn, does the called boundary move in the direction of the correction, does that response remain concentrated on the reviewed edge, and is it still detectable in the umpire's next game?'' Answering these questions requires comparisons both within the same game and across consecutive games.

\subsection{Method}
\label{sec:rq1-method}

\subsubsection{Identifying the Immediate Feedback Response}
\label{sec:rq1-method-design}

To identify the response associated specifically with corrective feedback, we compare overturned challenges with upheld challenges. While both expose the umpire to automated review, only an overturn indicates that the original call disagreed with the automated ruling. The contrast therefore separates responses following corrective feedback from those following an upheld review. It is not randomized, however, since overturned and upheld challenges may differ in pitch location and in the circumstances that led players to challenge. Accordingly, we interpret the contrast as a differential response following corrective versus non-corrective review rather than as an isolated causal effect of feedback.

Beyond asking whether calls change after an overturn, we also ask whether the response remains specific to the reviewed boundary. We compare subsequent pitches on the reviewed edge with those on the other three edges. A response concentrated on the reviewed edge would indicate a local adjustment, whereas movement on the other edges would suggest a broader change in the called zone. These two comparisons determine how challenge exposure is represented in the model.

For each pitch $i$, let $e_i$ denote its edge and $o_i$ its order within game $g_i$. We classify challenges by verdict $v \in \{\mathrm{ov},\mathrm{up}\}$ and by the direction implied by the original call $s \in \{\mathrm{tight},\mathrm{wide}\}$. The direction of the correction depends on the original call. A challenged ball corresponds to a potentially too-tight zone, while a challenged strike corresponds to a potentially too-wide zone. Let $F_{g,e}^{v,s}$ be the pitch order of the first challenge of type $(v,s)$ on edge $e$ in game $g$, with $F_{g,e}^{v,s}=\infty$ when no such event occurs. We then define

\begin{equation}
\label{eq:own-oth}
A_i^{\mathrm{own},v,s}
=
\mathbf{1}\!\left[o_i > F_{g_i,e_i}^{v,s}\right],
\qquad
A_i^{\mathrm{oth},v,s}
=
\mathbf{1}\!\left[o_i >
\min_{e \neq e_i} F_{g_i,e}^{v,s}\right].
\end{equation}

The first indicator marks pitches after a ruling of type $(v,s)$ on the same edge, while the second marks pitches after such a ruling on another edge. Once activated, each remains on for the rest of the game. The analysis therefore measures how calls change after the first relevant ruling of each type rather than treating every repeated challenge as a separate exposure.

With challenge exposure defined in this way, we first estimate changes in boundary position. Let $m \in \{\mathrm{own},\mathrm{oth}\}$ denote whether exposure occurred on the pitch's own edge or another edge. We fit

\begin{equation}
\label{eq:eta-alpha}
\eta_i
=
a_{c_i}
+
\beta_d d_i
+
\mathbf{x}_i^\top \boldsymbol{\gamma}
+
\sum_{m,v,s}
\tau_{m,v,s} A_i^{m,v,s},
\end{equation}

where $c_i$ is the game-by-edge cell, $d_i$ is the signed distance from the relevant reference edge defined in Section~\ref{sec:data-frame}, and $\mathbf{x}_i$ contains covariates that vary within a cell. The coefficient $\tau_{m,v,s}$ shifts the calling curve after the corresponding exposure begins, representing a change in where the called boundary lies.

We also examine whether a ruling changes the sharpness of the ball-strike transition. To do so, we allow challenge exposure to modify the distance slope

\begin{equation}
\label{eq:eta-sigma}
\eta_i
=
a_{c_i}
+
\beta_d d_i
+
\mathbf{x}_i^\top \boldsymbol{\gamma}
+
\sum_{m,v,s}
A_i^{m,v,s}
\left(
\tau_{m,v,s}
+
\delta_{m,v,s}d_i
\right).
\end{equation}

When an exposure is active, the distance slope becomes $\beta_d+\delta_{m,v,s}$. Thus, $\tau$, which shifts the calling curve, captures where the called boundary moves, while $\delta$, which changes the distance slope, captures whether calls around that boundary become sharper or more gradual.

\subsubsection{Controlling for Game-Level Conditions}
\label{sec:rq1-method-clogit}

To separate changes following a challenge ruling from conditions shared within the same game, we compare pitches within game-by-edge cells. Ballpark characteristics, environmental conditions, and other factors shared within a cell may affect umpire calls while also being related to when challenges occur. A conditional logistic model keeps the comparison within the same game and edge while accounting for these shared conditions.

An ordinary fixed-effect logistic model would require a separate intercept $a_c$ for every game-by-edge cell. Each cell contains fewer than eight calls on average, making direct estimation of thousands of cell-specific intercepts prone to small-cell estimation problems. Conditional logistic estimation instead conditions on the number of strike calls within each cell and removes $a_c$ from the likelihood. The exposure coefficients are therefore identified from variation among pitches on the same edge within the same game. Umpire indicators are also constant within these cells and drop out of the conditional likelihood.

Removing the cell intercept prevents us from recovering the absolute level of $\alpha$, which describes where the called boundary lies relative to the common reference. The event-level analysis does not require this absolute position. Instead, the exposure coefficients identify how far the called boundary moves after a ruling within the same game.

\subsubsection{Estimating Boundary-Specific Effects}
\label{sec:rq1-method-contrasts}

We next translate the exposure coefficients into the spatial contrasts reported in the results. For each edge membership $m$ and ruling direction $s$, we compare the post-ruling shift after an overturn with the corresponding shift after an upheld challenge

\begin{equation}
C_{m,s}
=
\tau_{m,\mathrm{ov},s}
-
\tau_{m,\mathrm{up},s}.
\end{equation}

Here, $C_{m,s}$ captures the additional shift in the calling curve following an overturn relative to an upheld review. Since this contrast remains on the logistic scale, we convert it to the distance scale from Section~\ref{sec:data-metrics}

\begin{equation}
\label{eq:contrasts}
\Delta_{\mathrm{own}}(s)
=
\frac{C_{\mathrm{own},s}}{\beta_d},
\qquad
\Delta_{\mathrm{targeted}}(s)
=
\frac{
C_{\mathrm{own},s}
-
C_{\mathrm{oth},s}
}{\beta_d}.
\end{equation}

The quantity $\Delta_{\mathrm{own}}$ measures how far the reviewed boundary moves after an overturn relative to an upheld challenge. The quantity $\Delta_{\mathrm{targeted}}$ compares that movement with the other three edges and therefore indicates how strongly the response is concentrated on the reviewed boundary. Under our sign convention, a positive $\Delta_{\mathrm{own}}$ means that the called boundary moves outward.

We apply the same overturn-versus-upheld comparison to transition width. Since $\sigma$ is the reciprocal of the distance slope, its value under each exposure state is

\begin{equation}
\label{eq:dsigma}
\sigma_{m,v,s}
=
\frac{1}{\beta_d+\delta_{m,v,s}},
\qquad
\Delta_{\sigma,m}(s)
=
\sigma_{m,\mathrm{ov},s}
-
\sigma_{m,\mathrm{up},s}.
\end{equation}

The quantity $\Delta_{\sigma,m}$ therefore measures whether the ball-strike transition becomes sharper or more gradual after an overturn relative to an upheld challenge. No additional conversion is needed since $\sigma$ is already expressed in inches. Standard errors for these nonlinear contrasts are computed using the delta method.

\subsubsection{Testing Persistence across Consecutive Games}
\label{sec:rq1-method-persistence}

The within-game analysis identifies the immediate response following a challenge ruling, but persistence requires examining the umpire's subsequent game. Since previous-game exposure is constant within the current game, a current-game fixed effect would absorb it. We therefore return to the model from Section~\ref{sec:data-metrics} with umpire indicators and restrict the sample to pitches before the first overturn of the new game. This opening window allows us to examine whether the post-ruling pattern remains detectable before any new corrective feedback is received.

Comparing these estimates with the within-game response provides a behavioral measure of how long the adjustment lasts. We use this comparison to characterize its time course without assigning the observed pattern to a particular cognitive mechanism.

\subsubsection{Samples}
\label{sec:rq1-method-samples}

The season-level analysis uses the full band sample described in Section~\ref{sec:data-frame}. Event-level analyses use 2026, the only season in which challenge rulings are observed. The within-game sample contains 59{,}076 pitches across 7{,}040 game-by-edge cells, while the next-game analysis contains 25{,}938 pitches from the opening windows. Standard errors for both event-level analyses are clustered by umpire.

\subsection{Results}
\label{sec:rq1-results}

\subsubsection{Season-Level Change under Reviewability}
\label{sec:rq1-zone}

\begin{figure*}[t]
  \centering
  \includegraphics[width=\linewidth]{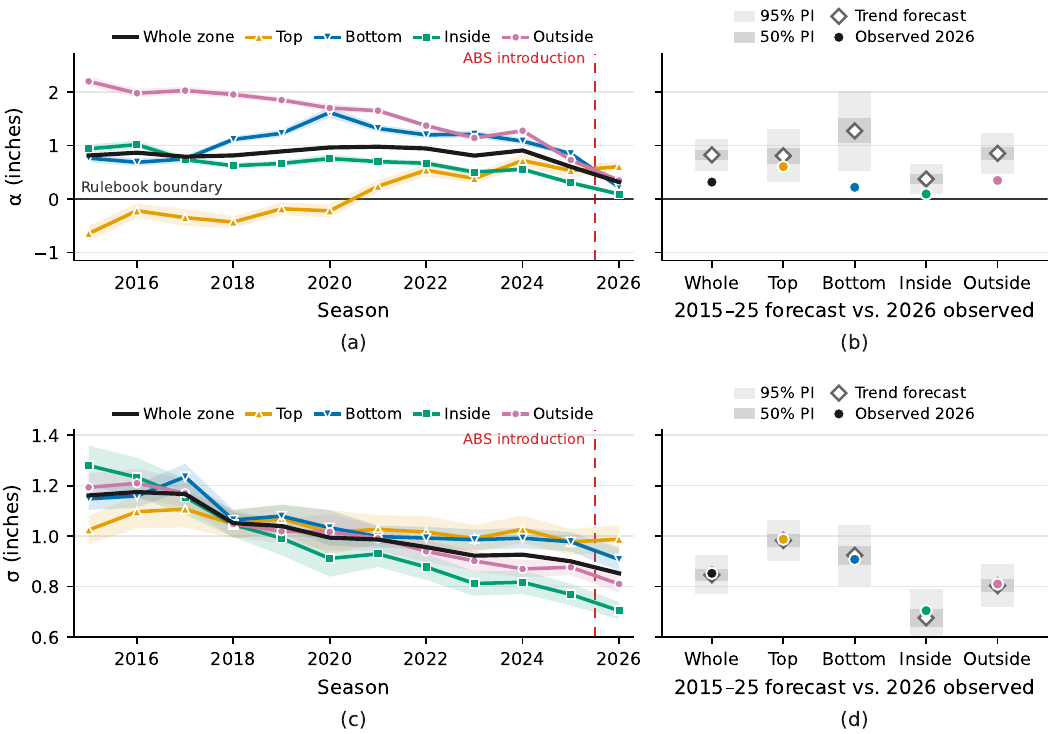}
  \caption{Seasonal changes in boundary position $\alpha$ and transition width $\sigma$ before and after challenge-based ABS. Figure~\ref{fig:alpha-trajectory} (a, c) show annual estimates from 2015--2026, with the dashed line marking ABS introduction. Figure~\ref{fig:alpha-trajectory} (b, d) compare the observed 2026 estimates ($\circ$) with forecasts extrapolated from the 2015--2025 trend ($\diamond$); shaded regions show the 50\% and 95\% prediction intervals. The 2026 boundary positions shift toward the common reference, particularly at the bottom and outside edges, whereas transition widths remain close to their pre-2026 trends.}
  \Description{Four panels compare annual estimates of boundary position and transition width with values predicted from the pre-adoption trend. The 2026 estimates show a clearer departure from the previous trajectory for boundary position than for transition width, with the largest edge-level departures in boundary position appearing on the bottom and outside edges.}
  \vspace{-3ex}
  \label{fig:alpha-trajectory}
\end{figure*}


We first ask whether challenge-based ABS coincided with a system-level change in umpire judgment beyond the changes already underway before 2026, focusing on boundary position $\alpha$ and transition width $\sigma$. A departure in $\alpha$ would indicate that the effective calling boundary moved, whereas a departure in $\sigma$ would indicate a change in boundary sharpness. Since both quantities had already evolved over preceding seasons, we compare the 2026 observations with forecasts from the 2015--2025 trajectory rather than treating 2025 as a stationary baseline.

Figure~\ref{fig:alpha-trajectory} shows that the two quantities behave differently in 2026. Panels (a) and (b) show the longer-term evolution and forecast comparison for $\alpha$. The whole-zone, bottom, inside, and outside estimates fall below their pre-adoption forecasts with $p<0.05$, while the top edge remains within its predicted range with $p>0.05$. The largest departures occur at the bottom and outside edges, where the called boundary had historically extended farther outside the common reference. In contrast, panels (c) and (d) show that the 2026 $\sigma$ estimates remain close to their forecasts across the whole zone and all four edges, with $p>0.05$ throughout.

\textbf{Taken together, these results indicate that umpire calls changed primarily in where the boundary was placed, while the sharpness of that boundary remained consistent with its pre-adoption trajectory.}

\subsubsection{Immediate Response to Challenge Feedback}
\label{sec:rq1-response}

\begin{figure*}[t]
  \centering
  \includegraphics[width=\linewidth]{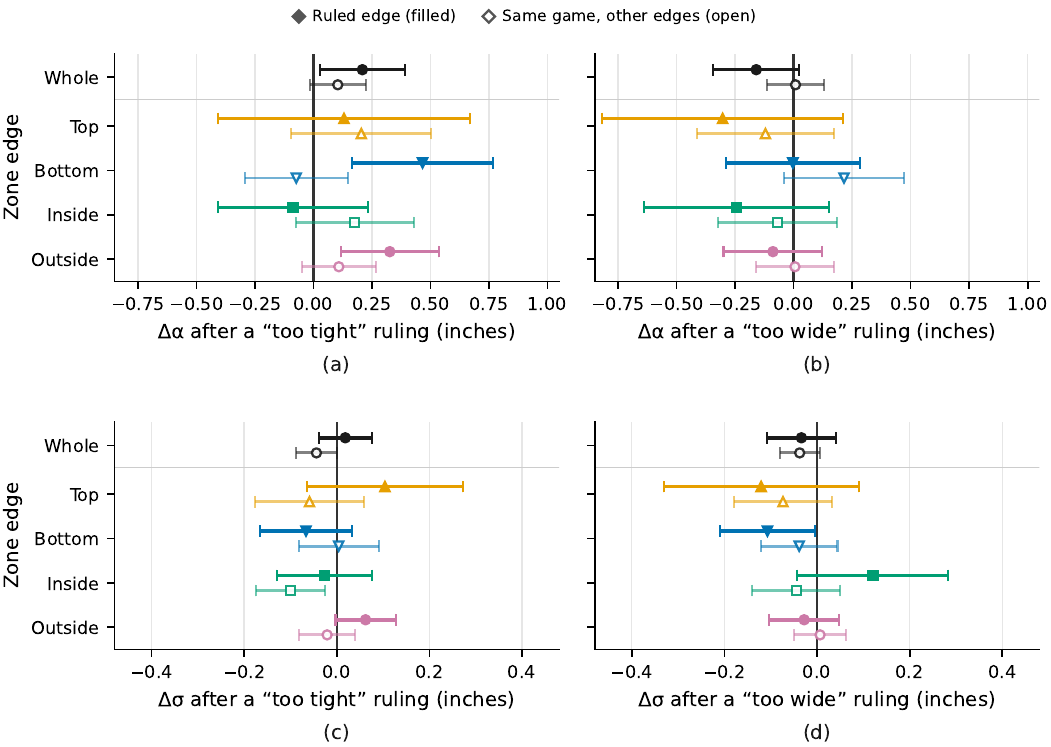}
  \caption{Within-game changes after challenge rulings. (a) and (b) show changes in boundary position $\Delta\alpha$; (c) and (d) show changes in transition width $\Delta\sigma$. Filled markers indicate changes on the reviewed edge, while open markers indicate changes on the other three edges in the same game; the horizontal bars show 95\% confidence intervals. After a ``too tight'' ruling, positive $\Delta\alpha$ corresponds to expanding the called zone, whereas after a ``too wide'' ruling, negative $\Delta\alpha$ corresponds to contracting it. Boundary-position changes are directionally concentrated on the reviewed edge, while changes in $\sigma$ show no consistent pattern.}
  \Description{Four dot-and-whisker panels compare post-ruling changes on the reviewed edge with changes on the other edges. Estimates on the other edges remain close to zero across most conditions. Boundary-position estimates on the reviewed edge show directionally different responses depending on the ruling, whereas transition-width estimates show no consistent direction. Confidence intervals are wide for many individual edge estimates.}
  \label{fig:edge-response}
  \vspace{-3ex}
  
\end{figure*}

The season-level analysis shows that boundary position departed from its previous trajectory in 2026, but it cannot reveal whether a particular automated correction is followed by a corresponding change in subsequent calls. At the event level, this leads to two expectations. A ball overturned to a strike indicates that the called boundary was too tight and should move outward, whereas a strike overturned to a ball indicates that it was too wide and should move inward. If this response is specific to the corrected spatial judgment, it should also be concentrated on the reviewed edge. We test these expectations by comparing overturned with upheld challenges and then the reviewed edge with the remaining edges. Since overturned and upheld challenges are not randomly assigned, we interpret these estimates as differential responses following corrective versus non-corrective review rather than as an isolated causal effect of feedback.

Figure~\ref{fig:edge-response} shows both the directional and spatial patterns in boundary position. In panel (a), reviewed-edge estimates generally move outward after a ``too tight'' ruling, while panel (b) shows a tendency toward inward movement after a ``too wide'' ruling. The outward pattern is most visible at the bottom and outside boundaries. Estimates for the other three edges remain much closer to zero under both ruling directions, and none reaches statistical significance with $p \geq 0.086$. Most reviewed-edge confidence intervals still include zero, so the data provide clearer evidence about the direction and spatial concentration of the response than its precise magnitude at any single edge. This uncertainty reflects the relatively few pitches within each game-by-edge cell.

Panels (c) and (d) examine whether the same rulings are followed by a change in boundary sharpness rather than boundary position. Here, $\Delta\sigma$ varies in sign across edges and ruling directions, and most confidence intervals overlap zero. We therefore find no consistent evidence that an overturn is followed by either a sharper or a more gradual ball-strike transition around the affected boundary.

\textbf{Taken together, these results indicate that the clearest immediate post-overturn pattern is a directionally appropriate change in boundary position localized to the reviewed edge, with no comparable evidence of spillover to the other edges or a systematic change in transition width.}

\subsubsection{Persistence across Consecutive Games}
\label{sec:rq1-persistence}

\begin{figure*}[t]
  \centering
  \includegraphics[width=\linewidth]{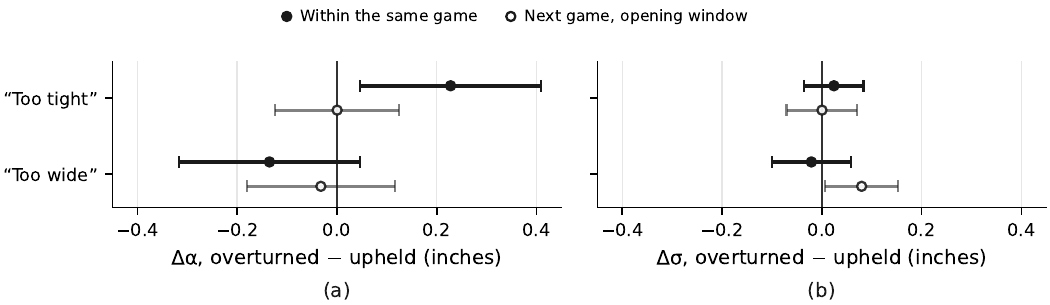}
  \vspace{-4ex}
  \caption{Persistence of post-challenge changes across games. (a) Changes in boundary position $\Delta\alpha$ and (b) changes in transition width $\Delta\sigma$. Filled markers indicate the remainder of the same game, while open markers indicate the opening window of the umpire's next game; the horizontal bars show 95\% confidence intervals. The directional change in boundary position observed within the same game largely disappears by the next game, while $\sigma$ remains close to zero at both time horizons.}
  \Description{Two dot-and-whisker panels compare within-game estimates with estimates from the opening window of the umpire's next game. The within-game boundary-position estimates differ by ruling direction, while the next-game estimates lie near zero. Transition-width estimates remain near zero on both time horizons.}
  \label{fig:carryover}
  \vspace{-3ex}
\end{figure*}

The preceding analysis shows that an overturn can be followed by a localized directional change within the same game, but RQ1 further asks whether that response persists beyond the immediate feedback context. A short-lived adjustment should be visible after the ruling but weaken by the next game, whereas a more persistent change should remain detectable when the umpire returns to call pitches later. We examine these alternatives by comparing the within-game estimates with those from the opening window of the umpire's next game, before another overturn can provide new corrective feedback.

Figure~\ref{fig:carryover} (a) shows a clear difference between these two time horizons for boundary position. Within the same game, a ``too tight'' ruling is followed by an outward movement of the called boundary with $p=0.014<0.05$. A ``too wide'' ruling is followed by movement in the opposite direction, although this estimate is less precise and does not reach statistical significance with $p=0.145>0.05$. In the opening window of the next game, however, neither directional response remains detectable. Both estimates lie near zero, with $p=0.996>0.05$ after a ``too tight'' ruling and $p=0.670>0.05$ after a ``too wide'' ruling. Panel (b) shows a similar lack of persistence in transition width, with $\Delta\sigma$ remaining close to zero at both time horizons.

The contrast across games is therefore more consistent with an adjustment that is strongest shortly after corrective feedback than with one that clearly persists into later judgments. This time course resembles the distinction in perceptual-learning and motor-learning research between rapidly adjustable responses and more persistent recalibration~\cite{robert1998goldstone,taylor2014explicit}, although the behavioral data do not identify the underlying mechanism. The next-game confidence intervals remain wide enough that smaller carryover effects cannot be ruled out. 

\textbf{The results show that the directional change in boundary position is detectable primarily within the game in which the correction occurs and is not detectably preserved at the beginning of the next game.}

\begin{keytakeaway}
\textbf{Key takeaway of RQ1.}
The clearest changes under challenge-based ABS appear in boundary position rather than transition width. Individual overturns are followed by localized directional adjustments within the same game, but these responses are not detectably preserved into the next game.
\end{keytakeaway}

\section{RQ2. How Did Systematic Variation in Umpire Calls Change Under Challenge-Based ABS?}
\label{sec:rq2}

\subsection{Analytical Overview}
\label{sec:rq2-question}

RQ1 examined how the called strike zone changed under challenge-based ABS and how umpires responded to individual overturns. Those broader changes need not imply that previously documented variation across conditions changed in the same way. We therefore ask whether systematic variation associated with ball-strike count, player status, and pitch type also changed in 2026.

For count and player status, we distinguish differences in boundary position $\alpha$ from differences in transition width $\sigma$. Unlike count and player status, pitch type is a property of the pitch itself and may affect how difficult it is to judge. We therefore characterize pitch-type differences using transition width and miscall rates rather than criterion bias. This distinction also carries into the historical comparison. Count and player status are evaluated against their pre-adoption trajectories, while pitch type is summarized against the 2015--2025 average, with statistical departures evaluated against the pre-adoption trend.

\vspace{-1ex}
\subsection{Method}
\label{sec:rq2-method}

\subsubsection{Condition-Specific Criterion and Transition Width}
\label{sec:rq2-method-model}

To measure how umpire calls vary across conditions, we extend the regression from Section~\ref{sec:data-metrics}. For a categorical factor with reference group $g_0$, let $G_{ig}$ indicate that pitch $i$ belongs to group $g$. Suppressing the adjustment terms already defined in Section~\ref{sec:data-metrics}, the model becomes

\begin{equation}
\label{eq:rq2-condition}
\eta_i
=
\eta_i^{(0)}
+
\sum_{g \neq g_0}
G_{ig}
\left(
\tau_g + \delta_g d_i
\right),
\end{equation}

where $\eta_i^{(0)}$ is the baseline linear predictor and $d_i$ is the signed distance from the relevant zone boundary. The coefficient $\tau_g$ shifts the calling curve relative to the reference group, capturing a difference in boundary position, while $\delta_g$ changes the distance slope and captures a difference in boundary sharpness.

We convert the shift associated with $\tau_g$ to the condition-specific criterion $\alpha_g$ using the transformation in Section~\ref{sec:data-metrics}. The corresponding transition width is

\vspace{-3ex}
\begin{equation}
\label{eq:rq2-sigma}
\sigma_g = \frac{1}{\beta_d + \delta_g}.
\end{equation}

This decomposition lets us distinguish two forms of systematic variation. $\alpha_g$ describes where the called boundary lies for condition $g$, while $\sigma_g$ describes how sharply calls change around that boundary.

\vspace{-1ex}
\subsubsection{Count and Player Status}
\label{sec:rq2-method-context}

Count is represented by the twelve possible ball-strike counts, with 0-0 as the reference category, and uses the full band sample from Section~\ref{sec:data-frame}. Since count is the focal factor in this analysis, the count terms included as adjustment covariates in the baseline model are omitted.

To represent career status using information established before each season, we use the cumulative number of All-Star selections. Moreover, different cut points are used by role to avoid sparse pitcher categories. Batters are grouped into no previous selections, one to three selections, and four or more selections, whereas pitchers are grouped into no previous selections, one selection, and two or more selections. Since each pitch involves both a batter and a pitcher, both status variables enter the model simultaneously. We treat these groups as indicators of career status rather than as causal effects of receiving an All-Star selection.

\vspace{-1ex}
\subsubsection{Pitch-Type Miscall Rates}
\label{sec:rq2-method-pitch}

We analyze six Statcast pitch types, including four-seam fastballs, sinkers, cutters, sliders, curveballs, and changeups, using four-seam fastballs as the reference category and retaining only cells with at least 209 pitches.
Transition width $\sigma$ describes how sharply calls change around the boundary, but it does not show the direction in which calls disagree with the reference zone. We therefore separate miscalls into two directions

\begin{equation}
\label{eq:rq2-miscall}
r_{\mathrm{wide}}
=
\frac{
N(\text{outside-zone pitch called strike})
}{
N(\text{outside-zone pitches})
},
\qquad
r_{\mathrm{tight}}
=
\frac{
N(\text{inside-zone pitch called ball})
}{
N(\text{inside-zone pitches})
}.
\end{equation}

Here, $r_{\mathrm{wide}}$ measures how often pitches outside the reference zone are called strikes, while $r_{\mathrm{tight}}$ measures how often pitches inside the zone are called balls. We also report the overall miscall rate, defined as the proportion of pitches in the analysis band for which the umpire's call disagrees with the reference zone, together with Wilson intervals.

Taken together, these measures help distinguish boundary movement from changes in boundary sharpness. Holding the pitch-location distribution fixed, an inward boundary shift should reduce wide-side errors while tending to increase tight-side errors. A sharper transition, by contrast, can reduce disagreement on both sides. The combination of $r_{\mathrm{wide}}$, $r_{\mathrm{tight}}$, and $\sigma$ therefore shows whether changes in miscall rates are more consistent with a shift in the called boundary or with sharper calls around it.


\subsection{Results}
\label{sec:rq2-results}

\begin{figure}[t]
  \centering
  \includegraphics[width=\linewidth]{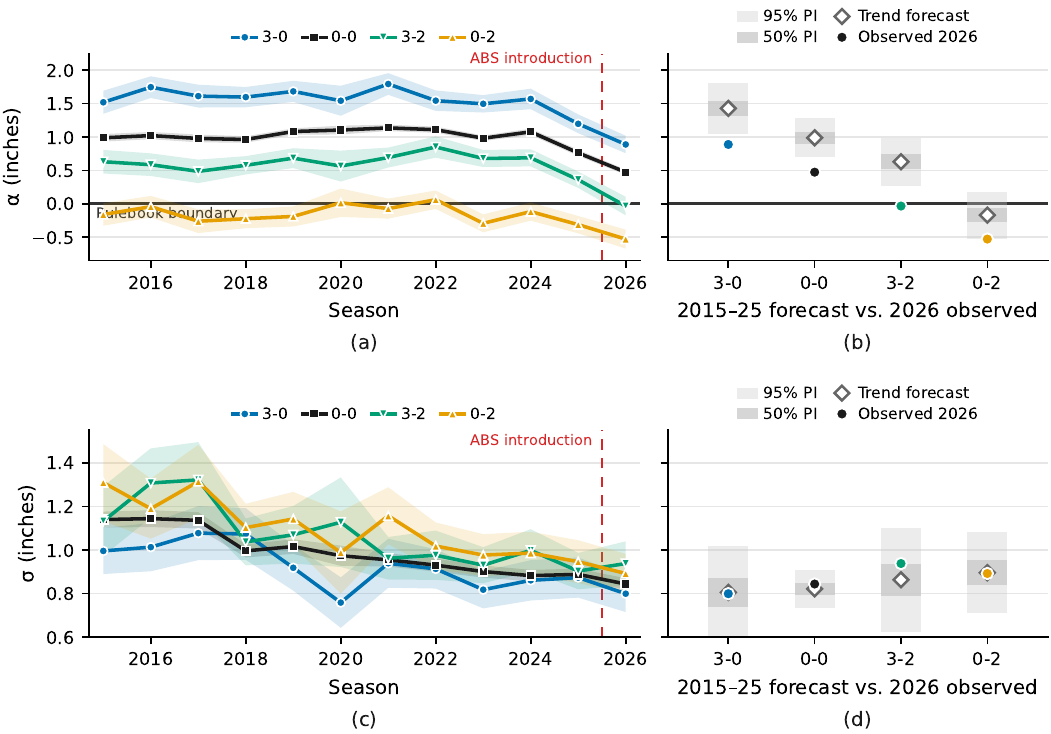}
  \vspace{-5ex}
  \caption{Count-dependent changes in boundary position $\alpha$ and transition width $\sigma$. (a, c) Seasonal estimates for four representative counts from 2015--2026. (b, d) Observed 2026 estimates compared with forecasts from the 2015--2025 trend, with 50\% and 95\% prediction intervals. In 2026, $\alpha$ shifts below the pre-adoption forecasts across all four counts, with several estimates falling outside the 95\% prediction intervals, while $\sigma$ remains comparatively stable.}
  \Description{Four plots compare count-dependent boundary position and transition width across seasons and against forecasts from the pre-adoption trend. In (a), alpha is shown from 2015 to 2026 for the 3-0, 0-0, 3-2, and 0-2 counts. Their ordering remains stable, with the widest called boundary at 3-0 and the tightest at 0-2. In (b), all four observed 2026 alpha estimates lie below their forecasted point estimates. The 3-0, 0-0, and 3-2 estimates fall below the corresponding 95 percent prediction intervals, while 0-2 lies near the lower bound. In (c), sigma shows smaller differences across counts. In (d), the observed 2026 sigma estimates remain close to their trend forecasts and within the corresponding prediction intervals.}
  \label{fig:count-bias}
  \vspace{-3ex}
  
\end{figure}

\subsubsection{Count-Dependent Calling}
\label{sec:rq2-count}

RQ1 showed that boundary position shifted inward in 2026, but this system-level change does not tell us whether established differences across ball-strike counts changed with it. We therefore ask whether count-dependent calling became less pronounced or whether the familiar ordering remained while the boundaries shifted together. We estimate boundary position $\alpha$, which captures where the called boundary lies, and transition width $\sigma$, our measure of boundary sharpness, across all 12 counts and compare the 2026 estimates with their 2015--2025 trajectories. Figure~\ref{fig:count-bias} shows four representative counts spanning the observed range.

Figure~\ref{fig:count-bias} (a) shows that the established ordering in boundary position remains visible in 2026. The called boundary is still widest at 3--0 and tightest at 0--2, with 0--0 and 3--2 between them. Against their pre-adoption trajectories, however, all four displayed counts shift inward. Panel (b) places the 3--0, 0--0, and 3--2 estimates beyond their 95\% prediction intervals, while 0--2 lies near the lower bound. The broader shift observed in RQ1 therefore extends across counts without removing their relative ordering.

A comparable departure does not appear in boundary sharpness. Figure~\ref{fig:count-bias} (c, d) shows the 2026 transition-width estimates $\sigma$ remaining close to their predicted trajectories, with $p>0.05$ for all four displayed counts. Since the counts begin from different boundary positions, the same inward movement has different consequences. At 3--0, where the called boundary had historically been wider than the common reference, the shift brings it closer to that reference, while at 0--2 it can move an already tighter boundary farther inside.

\begin{figure}[t!]
  \centering
  \includegraphics[width=\linewidth]{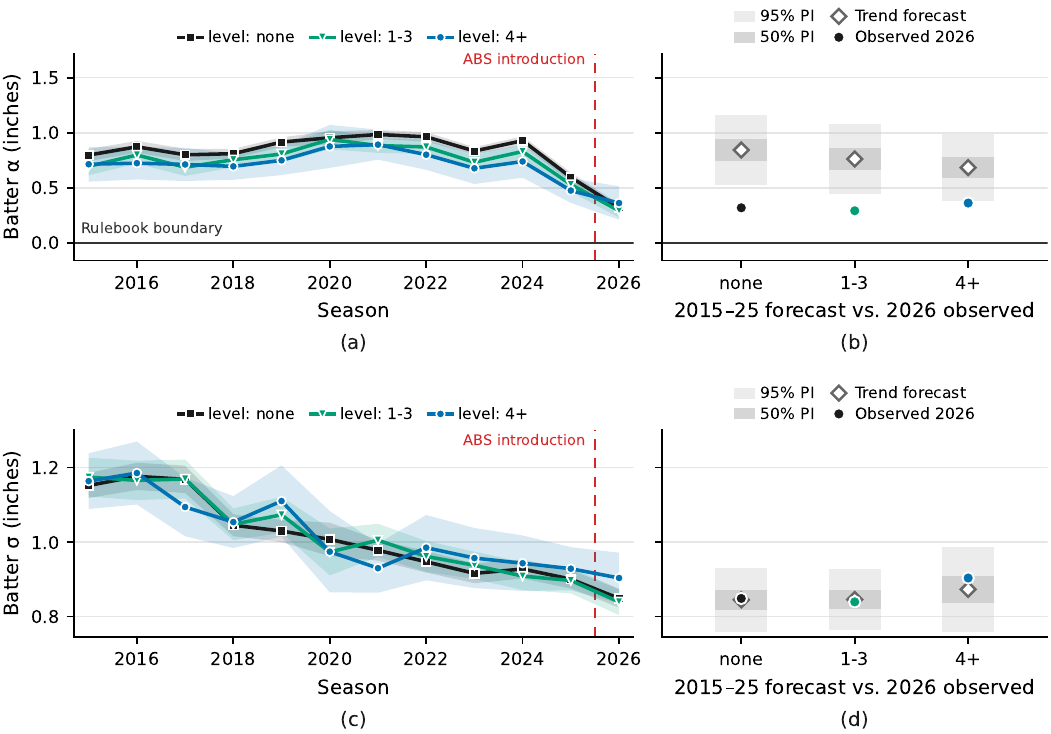}
  \vspace{-4.4ex}
  \caption{Boundary position $\alpha$ and transition width $\sigma$ by batter All-Star status. (a, c) Seasonal estimates for batters with no prior All-Star selections, one to three selections, and four or more selections. (b, d) Observed 2026 estimates compared with forecasts from the 2015--2025 pre-adoption trend, with 50\% and 95\% prediction intervals. In 2026, $\alpha$ falls below its pre-adoption forecast across all three status groups and the differences between groups narrow, while $\sigma$ remains close to its predicted values.}
  \Description{Four plots compare boundary position and transition width across batter All-Star status. In (a), alpha is shown for batters with no previous All-Star selections, one to three selections, and four or more selections. Before 2026, the higher-status groups generally have tighter called boundaries, while the three estimates become more similar in 2026. In (b), the observed 2026 alpha estimates are compared with forecasts from the pre-adoption trend. In (c), sigma substantially overlaps across the three groups over time. In (d), the observed 2026 sigma estimates remain close to their predicted values.}
  \label{fig:star-batter}
  \vspace{-3ex}
\end{figure}

\textbf{Taken together, these results show that the broader 2026 shift changes the absolute position of the called boundary across counts without removing the established count-dependent ordering, while boundary sharpness remains consistent with its pre-adoption trajectory.}

\subsubsection{Player Status}
\label{sec:rq2-status}

The count analysis shows that established differences across game situations remain visible in 2026, but player-status differences need not be equally stable. We therefore compare boundary position $\alpha$ and transition width $\sigma$ across batter and pitcher All-Star status groups. Before 2026, higher-status batters generally received tighter called boundaries, whereas higher-status pitchers tended to receive wider ones, although some group estimates overlap.

Figure~\ref{fig:star-batter} (a) shows that the batter-status groups move substantially closer together in 2026 after remaining separated through much of the pre-adoption period. Figure~\ref{fig:star-batter} (b) further shows that all three 2026 $\alpha$ estimates fall below their pre-adoption forecasts, with each departure significant at $p<0.05$. In contrast, Figure~\ref{fig:star-batter} (c, d) shows that $\sigma$ remains close to its predicted trajectory, with $p>0.05$ for all three groups. Thus, the narrowing occurs primarily in \emph{where} the boundary is placed rather than in \emph{how sharply} it is applied.

\begin{figure}[t]
  \centering
  \includegraphics[width=\linewidth]{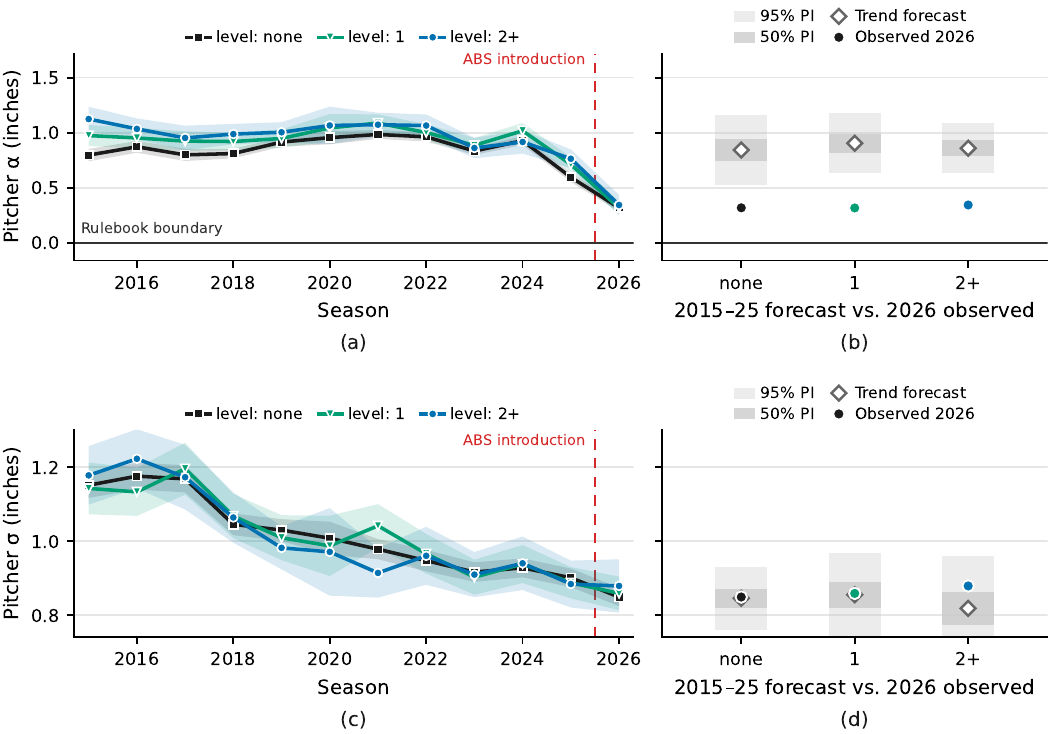}
  \caption{Boundary position $\alpha$ and transition width $\sigma$ by pitcher All-Star status. (a, c) Seasonal estimates for pitchers with no prior All-Star selections, one selection, and two or more selections. (b, d) Observed 2026 estimates compared with forecasts from the 2015--2025 pre-adoption trend, with 50\% and 95\% prediction intervals. In 2026, $\alpha$ falls below its pre-adoption forecast across all three status groups and the differences between groups narrow, while $\sigma$ remains close to its predicted values.}
  \Description{Four panels laid out as in the batter figure. Pitchers are grouped into no previous All-Star selections, one selection, and two or more selections. Before 2026, higher-status pitchers generally have wider called boundaries, although estimates for the highest-status group are imprecise. In 2026 the separation between the lower two groups narrows. Sigma estimates overlap substantially across status groups and remain within their forecast ranges.}
  \label{fig:star-pitcher}
\end{figure}

A similar pattern appears for pitchers. Figure~\ref{fig:star-pitcher} (a, b) shows that the three status groups converge to similarly lower boundary positions in 2026, with all observed $\alpha$ estimates well below their pre-adoption forecasts. Figure~\ref{fig:star-pitcher} (c, d) again shows little evidence of a status-specific change in $\sigma$, with $p>0.05$ across all three groups. Overall, challenge-based ABS coincides with a broad contraction of the called boundary and a narrowing of previously observed status differences, while leaving transition width comparatively unchanged.

\textbf{Taken together, these results show that status-associated differences narrow most clearly for batters in 2026, unlike the count-dependent ordering that remains clearly visible, while the evidence is weaker for pitchers and boundary sharpness remains broadly consistent with its pre-adoption trajectory.}

\subsubsection{Pitch-Type Differences}
\label{sec:rq2-pitch}

\newcommand{\stars}[1]{\makebox[0.9em][l]{\scriptsize #1}}

\begin{table*}[t]
  \caption{Miscall rates and transition width $\sigma$ by pitch type within the
  boundary band, comparing the 2015--2025 average with 2026. Stars mark a
  departure from the pre-adoption trend of Section~\ref{sec:data-trend}
  ($^{*}p<0.05$, $^{**}p<0.01$).}
  \label{tab:pitch-miscall}
  \centering\small
  \begin{tabular}{@{}llrrrrrr@{}}
    \toprule
    & & & \multicolumn{3}{c}{Miscall rate (\%)} & \multicolumn{2}{c}{$\sigma$ (in)} \\
    \cmidrule(lr){4-6} \cmidrule(l){7-8}
    Pitch & Type & $n$ & 2015--25 & 2026 & Relative change & 2015--25 & 2026 \\
    \midrule
    Four-seam fastball & all   & 319,326 & 27.1 & 19.9 & $-26.4$\stars{*}  & 1.00 & 0.82\stars{} \\
                       & wide  & 142,034 & 41.9 & 25.0 & $-40.3$\stars{**} &      &     \\
                       & tight & 177,292 & 15.3 & 15.4 & $+0.5$\stars{}    &      &     \\
    \addlinespace
    Sinker             & all   & 179,968 & 27.5 & 19.5 & $-29.0$\stars{**} & 1.05 & 0.83\stars{} \\
                       & wide  &  79,619 & 43.7 & 25.6 & $-41.5$\stars{**} &      &     \\
                       & tight & 100,349 & 14.7 & 14.5 & $-1.3$\stars{}    &      &     \\
    \addlinespace
    Cutter             & all   &  55,688 & 28.2 & 20.6 & $-26.8$\stars{}   & 1.05 & 0.82\stars{} \\
                       & wide  &  24,566 & 44.1 & 26.8 & $-39.2$\stars{**} &      &     \\
                       & tight &  31,122 & 15.5 & 14.8 & $-4.4$\stars{}    &      &     \\
    \addlinespace
    Slider             & all   & 106,188 & 27.7 & 21.0 & $-24.3$\stars{*}  & 1.03 & 0.87\stars{} \\
                       & wide  &  45,313 & 45.1 & 27.8 & $-38.4$\stars{**} &      &     \\
                       & tight &  60,875 & 14.9 & 15.1 & $+1.4$\stars{}    &      &     \\
    \addlinespace
    Curveball          & all   &  63,024 & 26.1 & 21.9 & $-16.2$\stars{}   & 1.00 & 0.94\stars{*} \\
                       & wide  &  25,729 & 43.5 & 29.2 & $-32.8$\stars{*}  &      &     \\
                       & tight &  37,295 & 14.3 & 15.5 & $+8.1$\stars{}    &      &     \\
    \addlinespace
    Changeup           & all   &  58,454 & 26.9 & 19.2 & $-28.6$\stars{**} & 0.95 & 0.85\stars{} \\
                       & wide  &  29,228 & 40.8 & 20.2 & $-50.5$\stars{**} &      &     \\
                       & tight &  29,226 & 13.3 & 18.1 & $+36.1$\stars{*}  &      &     \\
    \bottomrule
  \end{tabular}
\end{table*}

The count and status analyses ask whether established differences in boundary position persisted across game situations and players. Pitch type raises a different question. Since pitches may differ in how difficult they are to judge, we examine whether reduced disagreement with the reference zone appears across pitch types and which side of the boundary accounts for it. Table~\ref{tab:pitch-miscall} separates wide-side errors, where an outside pitch is called a strike, from tight-side errors, where an inside pitch is called a ball, together with transition width $\sigma$, our measure of boundary sharpness. An inward boundary shift should primarily reduce wide-side errors, whereas a sharper boundary should reduce errors on both sides.

Overall miscall rates are lower in 2026 for all six pitch types, with reductions from 16.2\% for curveballs to 29.0\% for sinkers. Relative to the pre-adoption trend, these reductions are significant for four-seam fastballs and sliders with $p<0.05$ and for sinkers and changeups with $p<0.01$. Cutters and curveballs also decline, but not significantly. The more consistent pattern appears on the wide side. For every pitch type, outside-zone pitches are substantially less likely to be called strikes in 2026, with reductions from 32.8\% to 50.5\%. All six depart significantly from their pre-adoption trends with $p<0.05$.

In contrast, tight-side errors do not show a common direction. They remain nearly unchanged for four-seam fastballs and sliders, decline slightly for sinkers and cutters, and increase for curveballs and changeups. Only the changeup increase is significant, rising by 36.1\% with $p<0.05$. This wide-side concentration is consistent with the inward boundary shift observed in RQ1, indicating that the overall decline in miscalls comes mainly from fewer outside-zone pitches being called strikes.

Boundary sharpness provides a complementary view of the pitch-type results. Although the 2026 transition-width estimates $\sigma$ are numerically lower than their 2015--2025 averages for all six pitch types, only the curveball departs significantly from its pre-adoption trajectory with $p<0.05$. Curveballs nevertheless retain the largest transition width in 2026 and the smallest descriptive reduction in overall miscalls. Prior work has associated curveballs with distinctive motion-perception effects~\cite{shapiro2010transitions}, which may help explain this residual difference, although the present analysis cannot distinguish that account from other pitch-specific factors.

\textbf{Taken together, these results show that the reduction in pitch-type miscalls is concentrated on wide-side errors across pitch types, which is more consistent with the broader inward shift in boundary position than with a uniform sharpening of calls around the boundary.}

\begin{keytakeaway}
\textbf{Key takeaway of RQ2.}
Systematic variation in umpire calls changes unevenly in 2026. Count-dependent ordering remains visible, status-associated differences narrow most clearly for batters, and pitch-type miscall reductions come mainly from fewer wide-side errors.

\end{keytakeaway}

\section{RQ3. How Do Players Decide When to Challenge?}
\label{sec:rq3}

\subsection{Analytical Overview}
\label{sec:rq3-question}

RQ1 and RQ2 examined how umpire calls changed under challenge-based ABS. We now turn to the other side of the system, where players decide which calls receive automated review. Since only challenged calls are reviewed, which errors ABS ultimately corrects depends partly on how players use their limited challenge opportunities.

To study this choice, we define a \emph{challenge opportunity} as a called pitch for which the eligible side still has at least one challenge remaining. Using these opportunities, we first ask whether challenge use increases as the geometric disagreement between the original call and the automated ruling grows. Exact pitch geometry is not available to players, however, which leads to a second question of whether challenge behavior aligns more closely with overturn probability based on that geometry or with probability inferred from information available at decision time. Finally, since an unsuccessful challenge reduces the team's remaining opportunities, we examine whether players become more selective when only one challenge remains.

\subsection{Method}
\label{sec:rq3-method}

\subsubsection{Challenge Opportunities and Geometric Error}
\label{sec:rq3-method-distance}

We first measure how far each pitch lies from a ruling that would favor the side eligible to challenge. Using the official zone that determines the ABS ruling, we compute the signed margin to the nearest boundary

\begin{equation}
\label{eq:cdist}
d_i
=
\min
\left\{
W' + x_i,\;
W' - x_i,\;
\mathrm{top}_i-z_i,\;
z_i-\mathrm{bot}_i
\right\},
\qquad
W'=W+\rho.
\end{equation}

The margin $d_i$ is positive inside the official zone and negative outside it. To put ball and strike calls on the same scale from the perspective of the eligible side, we orient this margin according to the original call

\begin{equation}
\label{eq:oriented}
m_i =
\begin{cases}
-d_i, & \text{if the original call was a strike},\\[2pt]
+d_i, & \text{if the original call was a ball}.
\end{cases}
\end{equation}

Under this convention, $m_i>0$ indicates a call that automated review would overturn in favor of the eligible side, while $m_i<0$ indicates a call that would be upheld. The magnitude $|m_i|$ gives the geometric distance from the relevant boundary.

Since $m_i$ is derived from tracked pitch location, it is available to the analyst but not directly to the player making the challenge decision. We therefore use it as a geometric benchmark for how challenge use changes with distance into or away from the overturnable region. We bin $m_i$ in 0.3-inch intervals over $[-4.2,4.2]$ inches and compute challenge rates separately for the batting and fielding sides, excluding bins with fewer than 50 opportunities. Challenge rates for $m_i>0$ therefore show how often geometrically overturnable calls receive review, although unchallenged calls do not by themselves indicate poor decisions because players do not observe the exact geometry represented by $m_i$.

\subsubsection{Geometry-Aware and Observable-Information Predictions}
\label{sec:rq3-method-models}

While the geometric benchmark identifies which calls favor an overturn, players must decide without observing that exact geometry. To distinguish overturn probability implied by exact geometry from what players can infer at decision time, we estimate two models. The \emph{geometry-aware model} includes the pitch's signed position relative to the automated zone
\[
\widehat{p}^{\,G}_i
=
P(\text{overturn}_i=1 \mid \text{geometry}),
\]

so $\widehat{p}^{\,G}_i$ represents overturn probability when exact spatial information is available. By contrast, the \emph{observable-information model} excludes exact zone-relative geometry and uses only information available when the challenge decision is made

\[
\widehat{p}^{\,O}_i
=
P(\text{overturn}_i=1 \mid \text{observable information}).
\]

Thus, $\widehat{p}^{\,O}_i$ represents overturn probability based on information that could inform the player's decision. We compare challenge rates against both predictions to ask which better tracks actual challenge use. Stronger alignment with $\widehat{p}^{\,O}_i$ would be consistent with an information constraint, although it would not establish perceptual uncertainty as the only reason that overturnable calls go unchallenged.

To avoid comparing behavior with predictions fitted to the same observations, both probabilities are generated out of sample using a walk-forward procedure. For each month, the models are trained only on earlier opportunities and then applied to the subsequent month.

\subsubsection{Remaining Challenge Inventory}
\label{sec:rq3-method-inventory}

Even when the available evidence favors an overturn, players may still conserve a challenge because an unsuccessful attempt reduces the team's remaining opportunities. We therefore ask whether the same predicted overturn probability leads to different challenge use depending on whether one or two challenges remain. Specifically, we compare challenge rates across predicted overturn probabilities for the two inventory states. A lower rate with one challenge remaining would indicate more conservative use of the final opportunity, whereas similar rates would suggest that remaining inventory plays a smaller role in the decision.

\subsubsection{Sample}
\label{sec:rq3-method-sample}
Across these analyses, the sample contains 146,725 challenge opportunities and 7,381 actual challenges, corresponding to an overall usage rate of 5.03\%.


\subsection{Results}
\label{sec:rq3-results}

\begin{figure*}[t]
  \centering
  \includegraphics[width=\linewidth]{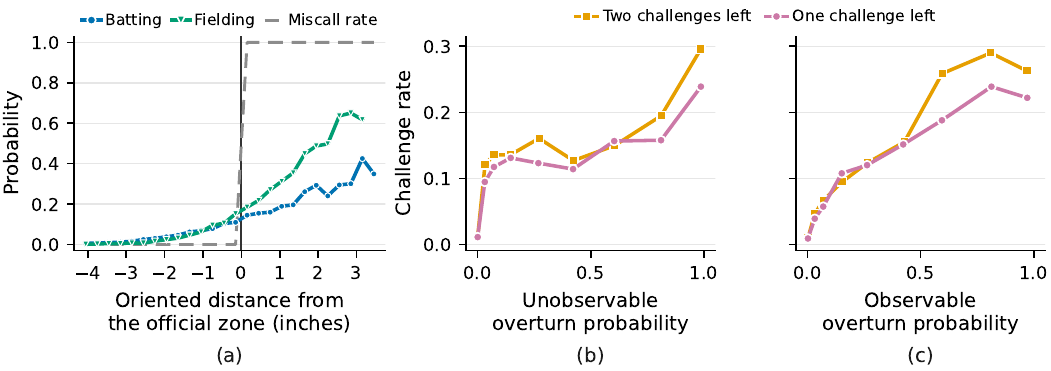}
  \caption{Challenge use as a function of call geometry and predicted overturn probability. (a) Challenge probability for the batting and fielding sides by oriented distance from the official zone; positive distance indicates calls that ABS would overturn, while the dashed line shows the corresponding miscall probability. (b, c) Challenge rates by predicted overturn probability using unobservable geometric information and observable information, respectively, separated by the number of challenges remaining. Challenge use rises more clearly with observable overturn probability, while having one versus two challenges remaining produces only modest differences.}
  \Description{Three plots characterize how players use challenges. In (a), challenge rate is plotted against oriented distance from the automated zone; positive values indicate calls that would be overturned. Challenge rates rise with oriented distance but remain below one even for clearly overturnable calls, and the fielding side challenges more often than the batting side. In (b), challenge rate is plotted against overturn probability predicted using exact pitch geometry and varies relatively little across much of the range. In (c), overturn probability is estimated using only observable information, and challenge rates increase more strongly with predicted success. Curves for one and two challenges remaining are similar over most of the range, with modest separation at higher predicted probabilities.}
  \label{fig:challenge-use}
\end{figure*}

\subsubsection{Challenge Rate by Geometric Error}
\label{sec:rq3-results-distance}

We first ask whether challenge rates rise with the geometric severity of calls that ABS would reverse. The oriented distance defined in Section~\ref{sec:rq3-method} places batting- and fielding-side opportunities on the same scale. Positive values indicate calls that automated review would overturn for the eligible side, and larger values indicate greater disagreement with the original call. If players respond to this disagreement, challenge rates should rise farther into the overturnable region.

Figure~\ref{fig:challenge-use}(a) shows this increase, but challenge rates remain well below one even when the geometry strongly favors an overturn. Between one and three inches into the overturnable region, batting-side rates range from approximately 0.19 to 0.30, while fielding-side rates range from approximately 0.31 to 0.65. Thus, many calls that ABS would overturn remain unchallenged despite an available opportunity to request review.

The fielding side also challenges more frequently than the batting side across much of the observed range. This separation may reflect differences in what the two sides can observe, when they consider a challenge, or the value of reversing the call. Since oriented distance captures the underlying geometry rather than information available to the player, this comparison cannot distinguish among these explanations.

\textbf{Taken together, these results show that challenge use increases with the geometric severity of an overturnable call, but many calls that ABS would reverse still receive no challenge.}

\subsubsection{Challenge Decisions by Available Information}
\label{sec:rq3-results-information}

The geometric pattern leaves an important question unresolved: exact pitch geometry determines whether ABS will overturn a call, but players do not observe it directly. We therefore compare two estimates of overturn probability. The geometry-aware probability $\widehat{p}^{\,G}$ uses the pitch's exact position relative to the automated zone, whereas the observable-information probability $\widehat{p}^{\,O}$ uses only information available at decision time. This comparison tests whether challenge decisions more closely follow the underlying geometry used by ABS or the information available to players.

Figure~\ref{fig:challenge-use} (b) shows only a weak relationship between challenge use and $\widehat{p}^{\,G}$. Across much of the middle range, substantially different geometry-based overturn probabilities correspond to similar challenge rates, spanning only about 0.06. By contrast, Figure~\ref{fig:challenge-use} (c) shows a much clearer increase with $\widehat{p}^{\,O}$, with challenge rates spanning approximately 0.21 over the corresponding range. Challenge rates nevertheless remain well below one even when $\widehat{p}^{\,O}$ is high, indicating that observable information alone does not explain every unchallenged error.

\textbf{Taken together, these results indicate that challenge decisions track information available to players at decision time more closely than the exact geometry used by ABS, although observable information does not fully explain why overturnable calls go unchallenged.}

\subsubsection{Challenge Decisions by Remaining Inventory}
\label{sec:rq3-results-inventory}

Even when observable evidence favors an overturn, players may preserve a challenge because an unsuccessful attempt reduces future review opportunities. We therefore ask whether remaining inventory changes how players respond to the same observable evidence. Holding $\widehat{p}^{\,O}$ fixed, stronger conservation would appear as lower challenge rates when only one challenge remains.

Figure~\ref{fig:challenge-use} (c) shows little separation between the two inventory states across most of the probability range. At low and moderate values of $\widehat{p}^{\,O}$, teams with one and two challenges remaining challenge at similar rates. The curves separate more clearly only at the upper end, where teams with two challenges remaining challenge somewhat more often. This pattern is consistent with modest conservation of the final challenge, but the inventory effect is small relative to the overall increase in challenge use with observable overturn probability.

\textbf{Taken together, these results suggest that remaining challenge inventory plays a secondary role in challenge use, with clearer inventory differences appearing mainly when an overturn already appears likely.}

\begin{keytakeaway}
\textbf{Key takeaway of RQ3.}
Players' challenge decisions align more closely with information available at decision time than with exact ABS geometry. Many overturnable calls remain unchallenged, while remaining challenge inventory has a comparatively smaller relationship with challenge use.
\end{keytakeaway}

\section{Discussion}
\label{sec:discussion}

\subsection{Real-Time Feedback and Player Calibration}
\label{sec:disc-feedback}

\begin{figure*}[t]
  \centering
  \includegraphics[width=\linewidth]{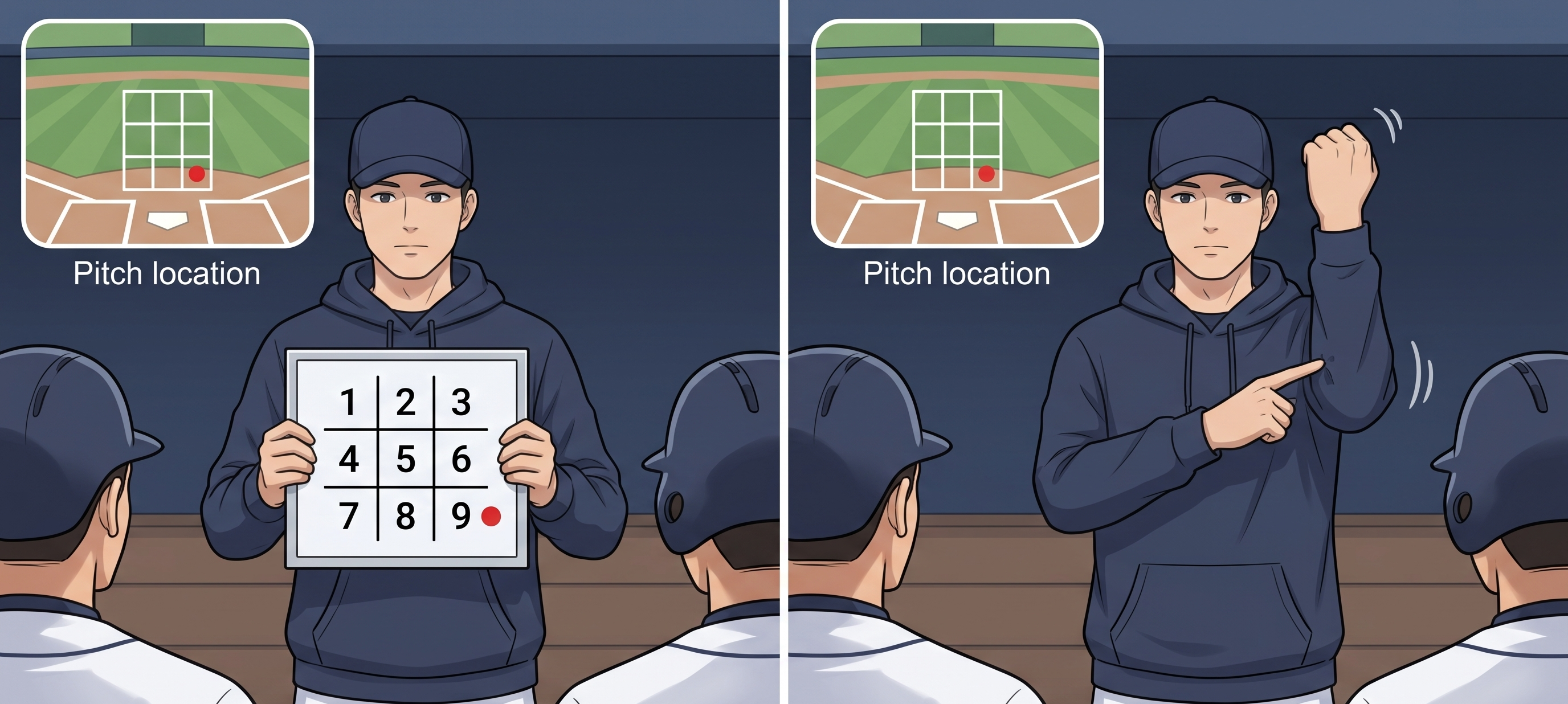}
  \caption{Illustrative examples of relaying ABS pitch-location information from the dugout. The numbered $3\times3$ zone board (left) and compact hand signal (right) translate pitch-location information into cues that players can interpret during play. These examples illustrate possible team-created feedback.}
  \Description{Examples of in-game communication of ABS pitch-location information in the KBO. On the left, a broadcast graphic shows a pitch location relative to a gridded strike zone, while a person in the dugout holds up a corresponding three-by-three numbered board with one region marked. On the right, a player uses a hand gesture that appears to convey pitch-location information. Faces are obscured.}
  \label{fig:kbo-feedback}
\end{figure*}

Our results suggest that access to information matters on both sides of the challenge system. In Section~\ref{sec:rq1-response}, an overturn was followed by a short-lived change in an umpire's subsequent calls. In RQ3, challenge decisions were more closely associated with information available to players than with the exact pitch geometry used by ABS. Together, these findings raise a broader question: what happens when automated officiating information becomes available to participants during play?

The KBO provides illustrative examples of how such feedback can emerge around full ABS. Because the system produces a ruling on every pitch, teams can potentially use its output as a continuing source of information about the automated zone. Figure~\ref{fig:kbo-feedback} illustrates two communication practices observed in actual KBO games and redrawn here for clarity: a numbered $3\times3$ grid identifying a region of the zone and a compact hand signal conveying pitch location. These examples are not intended as standardized KBO practices; rather, they show how teams can build their own communication layer around automated officiating.

This possibility is relevant to MLB's challenge format. Our RQ3 results show that many overturnable calls still go unchallenged and that challenge use more closely tracks information players can infer than exact pitch geometry. Additional feedback about previous pitches could therefore help players calibrate their estimate of the ABS boundary, the umpire's departures from it, and which borderline calls are worth contesting.

MLB's challenge system, however, provides a much thinner feedback stream than full ABS: only contested pitches reveal an ABS ruling, and those pitches are already selected because someone suspected an error. The resulting feedback is therefore sparse and potentially unrepresentative. This raises a broader Sports HCI design question: how much algorithmic feedback should participants receive, in what form, and at what point in play? Figure~\ref{fig:kbo-feedback} illustrates that when system output is available, participants may create such feedback channels themselves.

\subsection{Fans as Stakeholders in Automated Officiating}
\label{sec:fans}
 
Our analysis has been confined to the two parties who act inside the challenge system, the umpire whose call is reviewed and the player who requests the review. When a challenge is raised, however, the output of the system reaches not only these two but also the people watching the game, whether in the ballpark or on a broadcast. Fans neither make the call nor request the review, yet they are party to the process, and what it leaves them with is an experience of watching. What, then, has this system added to that experience?
 
To get a sense of this, we asked 32 viewers of a league in which the system rules on every pitch. What the challenge arrangement adds to the experience of watching is more visible to viewers who have watched baseball without challenges. They named two merits of the arrangement. The first is that established skills such as catcher framing---the catcher's technique of receiving borderline pitches in ways that increase the chance of a strike call---retain their value. The second is watching the new ability of players to judge whether to contest a call. This suggests that when a change to the rules opens a new angle on player ability, fans may take that angle up as a source of enjoyment.

These responses also suggest that fans should be considered as an important stakeholder in the design of challenge-based officiating systems. The information shown to viewers---what is revealed, when it is revealed, and how it is represented---may shape how they interpret player judgment, uncertainty, and the role of automation in the game. For example, selectively exposing pitch-location or overturn information could create new forms of anticipation around whether a player chooses to challenge, while other designs may better preserve uncertainty or highlight traditional baseball skills. This makes spectators not merely observers of the system, but stakeholders whose experience depends on how algorithmic information is disclosed. Future work should examine how the content, timing, and representation of such feedback shape spectators' understanding, enjoyment, and trust in automated officiating.
 
 
The survey items, respondent characteristics, and full response distributions are given in the supplementary material.

\subsection{Beyond the Ball Park}
While our study is grounded in professional baseball, challenge-based ABS illustrates a broader form of human--AI interaction in which automated judgment is selectively introduced into an otherwise human decision process. The umpire remains responsible for the initial call, while players decide when algorithmic review should be invoked. This creates two distinct interactions with automation: humans act under the possibility of later review, and others decide when that review is worth requesting.

Our findings suggest that the effects of such systems extend beyond whether the automated decision is more accurate. Review provides localized feedback about human judgment, but because it is selectively invoked, participants see only a subset of human--AI disagreements. What they learn from automation therefore depends not only on model outputs, but also on when those outputs become visible and who can request them.

This perspective applies to other human--AI systems that preserve human authority while providing selective algorithmic oversight. Designers may therefore need to consider the frequency, timing, and granularity of automated feedback alongside predictive accuracy. Challenge-based ABS provides a clear example of how these choices can shape both the person being reviewed and those who decide when automation enters the interaction.

\section{Conclusion}
\label{sec:conclusion}

Challenge-based ABS creates a distinctive form of selective human--AI review because umpires make every initial call, while players determine when ABS can intervene and issue the final ruling. This structure allows us to examine not only how umpire judgment changes under automated review, but also how players decide which judgments receive that review. Across twelve MLB seasons, we find that changes in umpire calls are most evident in boundary position rather than transition width. Individual overturns are followed by localized directional adjustments within the same game, although these responses are not detectably preserved into the next game. Systematic variation also changes unevenly, with count-dependent ordering remaining visible, status-associated differences narrowing most clearly for batters, and pitch-type miscall reductions coming mainly from fewer wide-side errors. On the player side, many overturnable calls remain unchallenged, and challenge decisions align more closely with information available at decision time than with exact pitch geometry. Together, these findings show that selective automated review depends both on how people respond when automation corrects them and on whether other participants choose to invoke that correction in the first place.
\section*{Ethics and Privacy Statement}

This work draws on two public sources and one small survey. Pitch-level tracking data come from the Statcast records that MLB publishes through Baseball Savant, and items absent from those records, namely umpire assignments, challenge rulings, and All-Star selections, were supplemented from MLB's public Stats API. No privileged access or non-public information was used. The umpires and players in the data act in professional capacities, and the calls in question are already public through broadcasts and official records, so no question of personal privacy arises. Because the analysis nonetheless concerns individual patterns of judgment, umpires are handled only as anonymous identifiers and every quantity we report is an estimate across the population of umpires. Observation covers part of a single season under a newly introduced system, so these results cannot serve as an assessment of any individual's competence.

The survey reported in the supplementary material was determined to be exempt from review by our institutional review board. Thirty-two adults completed it online, recruited by convenience from the personal networks of the research team. Participation was voluntary and uncompensated, respondents were informed of the purpose of the survey and told they could stop at any time, and no identifying information was collected.


\bibliographystyle{ACM-Reference-Format}
\bibliography{references}


\end{document}